\documentclass[12pt]{article}

\usepackage[utf8]{inputenc}
\usepackage[T1]{fontenc}
\usepackage{textcomp}
\usepackage{gensymb}
\usepackage{amsmath}
\usepackage[margin=1in]{geometry}
\usepackage{graphicx}
\usepackage{amssymb}
\usepackage{authblk}
\usepackage{blindtext}
\usepackage{apacite}
\allowdisplaybreaks
\usepackage{graphics}
\usepackage{float}
\usepackage{setspace}
\usepackage{amsthm}
\theoremstyle{plain}

\usepackage{algorithm}% http://ctan.org/pkg/algorithms
\usepackage{algpseudocode}% http://ctan.org/pkg/algorithmicx

\newcommand{\btheta}{ \mbox{\boldmath $ \theta $} }

\newcommand{\brho}{ \mbox{\boldmath $\rho$} }

\newcommand{\bz}{\textbf{z}}

\newcommand{\bK}{\textbf{K}}

\newcommand{\bs}{\textbf{s}}

\restylefloat{table}
\usepackage{subcaption}
\usepackage{amssymb}
\usepackage{hhline}
\RequirePackage[OT1]{fontenc}
\RequirePackage{amsthm,amsmath}
\RequirePackage{natbib}
\RequirePackage[colorlinks,citecolor=blue,urlcolor=blue]{hyperref}

\begin{document}

\title{Spatial Interpolation of Curves with Monotonicity Constraints: An Analysis of Antarctic Snow Density}
\author[1]{Philip A. White \thanks{Corresponding Author: pwhite@stat.byu.edu}}
\author[2]{Durban Keeler \thanks{durban.keeler@gmail.com}}
%\author[1]{William F. Christensen \thanks{william@stat.byu.edu}}
\author[2]{ Summer Rupper \thanks{ summer.rupper@geog.utah.edu}}
% \thankstext{t2}{First supporter of the project}
%\thankstext{t3}{Second supporter of the project}
%\runauthor{P. White et al.}
\affil[1]{Department of Statistics, Brigham Young University, USA}
\affil[2]{Department of Geography, University of Utah, USA}

\maketitle

\begin{abstract}

Snow density estimates below the surface, used with airplane-acquired ice-penetrating radar measurements, give a site-specific history of snow water accumulation. Because it is infeasible to drill snow cores across all of Antarctica to measure snow density and because it is critical to understand how climatic changes are affecting the world's largest freshwater reservoir, we develop methods that enable snow density estimation with uncertainty in regions where snow cores have not been drilled.   

In inland West Antarctica, snow density increases monotonically as a function of depth, except for possible micro-scale variability or measurement error, and it cannot exceed the density of ice. We present a novel class of integrated spatial process models that allow interpolation of monotone snow density curves. For computational feasibility, we construct the space-depth process through kernel convolutions of log-Gaussian spatial processes. We discuss model comparison, model fitting, and prediction. Using this model, we extend estimates of snow density beyond the depth of the original core and estimate snow density curves where snow cores have not been drilled. Along flight lines with ice-penetrating radar, we use interpolated snow density curves to estimate recent water accumulation and find predominantly decreasing water accumulation over recent decades.

\end{abstract}

% \begin{keyword}
% \kwd{sample}
% \kwd{\LaTeXe}
% \end{keyword}

\section{Introduction}\label{sec:intro}

Antarctic snow density is directly linked with climate drivers and ice sheet dynamics. Snow density measurements are also used in combination with airplane radar measurements to estimate surface mass balance (SMB) over time \citep{medley_constraining_2014}. As defined here, SMB is the net precipitation, sublimation, melt, refreeze, and wind redistribution of snow and is directly linked to changes in climate. More accurate quantification of SMB greatly improves our understanding of net mass balance processes, provides a direct link to climate drivers of ice sheet mass balance and ice sheet dynamics, and gives a reasonable target for climate and ice sheet process models. Because radar estimates of SMB require snow density measurements, accurate snow density estimation is essential. For this reason, researchers drill and analyze snow cores to measure snow density as a function of depth below the surface. However, snow cores often do not align with airplane flight lines with radar measurements. 

Because the density of ice, which we call $\rho_I$, is 0.917 g/cm$^3$, snow density can only take values between 0 g/cm$^3$ and $\rho_I$. Moreover, below the surface, snow density generally increases as a function of depth until it approaches the density of ice. Our goal here is to provide methods for estimating the snow depth-density curve in locations without drilled snow/ice cores while imposing appropriate functional constraints on our estimates of snow density. In this analysis, we develop models for Antarctic snow density as a function of depth below the surface that allow for spatial interpolation. Using these models, we predict snow density against depth in locations without data. Within this framework that shares information from neighboring cores, we extend snow density estimates to deeper depths than were originally drilled. 

There is a rich literature on monotonic or isotonic function estimation or regression \citep[see][for early discussion]{barlow1972,robertson1988}. This topic has been furthered under various modeling frameworks. In the frequentist literature, these examples include splines \citep{ramsay1988} and restricted kernels \citep{muller1988}. In a Bayesian framework, \cite{gelfand1991} and \cite{neelon2004} impose monotonicity through the prior distribution. More recently, \cite{riihimaki2010} propose monotonic curve estimation through constrained Gaussian processes (GP). Similarly, \cite{lin2014} use Gaussian process projections to estimate monotone curves.

Building on differential equation models for snow densification, we propose a class of novel monotone spatial processes constructed by integrating over positive space-depth processes with respect to depth, yielding a spatial process of monotone increasing functions. We call this class of models monotone integrated spatial processes (MISPs). Given the size and attributes of our dataset (Discussed in Section \ref{sec:data}), we choose to construct the positive space-depth process through kernel convolutions of log-Gaussian spatial processes \citep[see][for early discussion on process convolution models for spatial modeling]{higdon1998,higdon2002}. We compare the performance of various kernels, including Gaussian, $t$, asymmetric Laplace, and M-spline basis functions \citep{ramsay1988,meyer2008}. As long as the kernel is amenable to integration, we can represent the model as a constrained spatially varying coefficient model with integrated kernels as predictors \citep{gelfand2003}. This model provides simple spatial prediction of snow density curves by model-based interpolation of basis function coefficients.

We highlight three additional contributions motivated by this dataset. Because snow density can only take values between 0  g/cm$^3$ and $\rho_I$, we model these spatially-varying monotone curves in a unique hierarchical generalized linear model framework that corresponds to commonly used constrained differential equations for snow densification \citep[see, e.g.,][]{ herron_firn_1980, horhold_densification_2011, verjans2020bayesian}. In contrast to these common approaches, our approach allows this density curve to change its attributes to vary spatially. Snow density measurements are expected to be heteroscedastic because density measurements are taken as an average over some length of the core and are thus more certain when averaged over greater lengths. Moreover, the data come from four separate field campaigns, and, although their measurement methods are similar, we expect some differences in the measurement error associated with each group.
To account for these patterns in variability, we consider models for the variance that account for the length of the core used for the measurement and allow model uncertainty to depend on the group that took the measurement. Lastly, we apply the approach of \cite{keeler2020} to estimate SMB as a function of time using snow density curve predictions at locations with airplane acquired radar measurements, providing a much richer recent history of water accumulation over the Antarctic ice sheets. 
 
We continue this manuscript by discussing snow densification, including the differential equation framework that we adapt for our model, and its relationship with surface mass balance in Section \ref{sec:firn}. We explore the snow density dataset that motivates our statistical contributions in Section \ref{sec:data} and comment briefly on the ice-penetrating radar data used to estimate SMB. Then, we present the class of integrated spatial process models in Section \ref{sec:models}. Following our proposed methods, we give our final model for Antarctic snow density, including a discussion of our model comparison approach, model fitting, and spatial interpolation in Section \ref{sec:mod}. We then analyze the results of our model in Section \ref{sec:res} and conclude our paper with final comments and a discussion of possible extensions in Section \ref{sec:conc}.
 
\section{Snow Density: Physics and Data}\label{sec:Data_and_physics}

\subsection{Snow Densification and Surface Mass Balance}\label{sec:firn}
 
Previous efforts to model the variation of snow density with depth take various forms, building upon the physics of densification \citep{cuffey2010}, empirical fits to data using exponential functions \citep{miege_southeast_2013}, or a combination of the two \citep{herron_firn_1980, horhold_densification_2011, verjans2020bayesian,white2020modeling}. The hybrid approach by \cite{herron_firn_1980} is the most widely used model. Changes in density $\rho$ over depth $x$ are generally modeled using a special case Bernoulli's differential equation, 
\begin{equation}\label{eq:diff_eq}
    \frac{d}{dx} \log\left( \frac{\rho}{ \rho_I - \rho} \right) = z(x) 
\end{equation}
with solution, using integrating factors,
\begin{equation}\label{eq:sol_diff_eq}
\begin{aligned}
   \log\left( \frac{\rho(x)}{ \rho_I - \rho(x)} \right) &= \alpha + \int_0^x z(t) dt, \\
      \rho(x) &= \rho_I \frac{e^{\alpha + \int_0^x z(t) dt}}{1 + e^{\alpha + \int_0^x z(t) dt}} , 
      \end{aligned}
\end{equation}
where $\alpha$ is a constant determined by the density at depth $0$ and $z(x)$ is a positive function that varies over depth. In most literature \citep[see, e.g.,][]{herron_firn_1980,horhold_densification_2011,verjans2020bayesian}, $z(x)$ is modeled as a piece-wise constant function. To represent the positive function of depth $z(x)$ more generally, we consider kernel convolutions of log-Gaussian spatial processes (LGPs) in Section \ref{sec:models}. 

Ice sheet surface mass balance (SMB) refers to the net sum of all changes in mass added to an ice sheet's surface within a given year, encompassing solid precipitation, melting/refreezing snow, blown snow, and sublimation processes.
Due to below freezing temperatures year-round and the relatively small fraction of sublimation in most regions \citep{lenaerts_new_2012}, SMB in West Antarctica is reasonably approximated simply with falling and wind-blown snow.
The classical method of measuring SMB in remote and extreme environments like Antarctica involves collecting ice cores, determining an age-depth scale using seasonal markers in snow/ice physical properties and chemistry, and integrating snow/ice density over annual intervals to determine the total mass input (typically in water-equivalent depth) for a given year.
These methods present some limitations, with low spatial coverage being one of the most important.
Due to the extreme environments present in Antarctica, its lack of access, and the expense of data collection, relatively few annually resolved records of SMB exist in West Antarctica, with frequent clustering of coring sites.
This sparsity, combined with the fact that point measurements at times are representative of only a few square kilometers \citep{eisen_ground-based_2008, banta_spatial_2008}, limits the applicability of conclusions drawn from ice core studies.

An important advancement to address this limitation is the use of ice-penetrating radar surveys to image internal layering in the snow subsurface, providing much-needed spatially distributed coverage of SMB estimates \citep{koenig_remote_2014}.
%\textbf{If we wanted to include a figure of a radar echogram, this would probably be a good place for it.}
This method, however, is unable to provide information about the depth-density relationship of the snow, and snow density information is needed to produce SMB estimates. This necessitates independent estimates of density to expand the utility of SMB radar methods.

We apply approaches from \cite{keeler2020} to estimate surface mass balance in central West Antarctica over recent decades using estimates of snow density to 40 meters below the surface. This approach consists of computer vision algorithms, principally based on Radon transforms and peak finding, to pick annual snow layers in radar images to estimate annual SMB.
The method first performs Radon transforms on iterative local subsections of an input echogram image, followed by a subroutine to identify peaks in the radar return strength of individual radar traces (typically corresponding to annual snow layers).
These peaks are then grouped into laterally-continuous layers based on the integrated angle brightnesses determined in the Radon transform step and similarities in peak position, width, and magnitude.
Subsequent layers are assigned probabilities of representing annual layers based on layer length and return brightness.
Monte Carlo simulations using these probabilities, combined with the depth-density profiles generated using the methods outlined in this paper, produce individual annual SMB distributions for each trace location in the radar image.

\subsection{Data}\label{sec:data}

\begin{figure}[ht]
\vspace{-3mm}
\begin{center}
\includegraphics[width=0.49\textwidth]{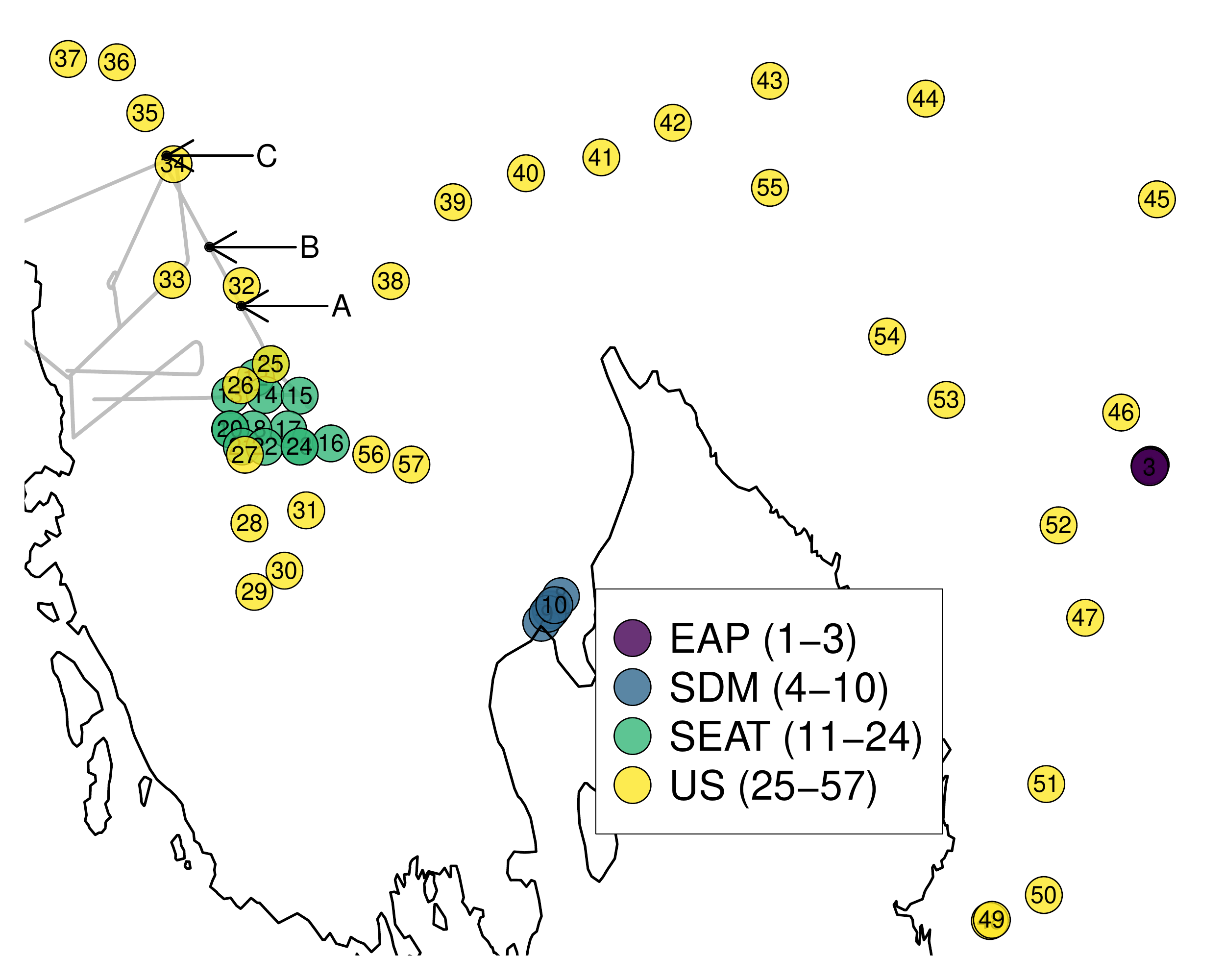}
\includegraphics[width=0.49\textwidth]{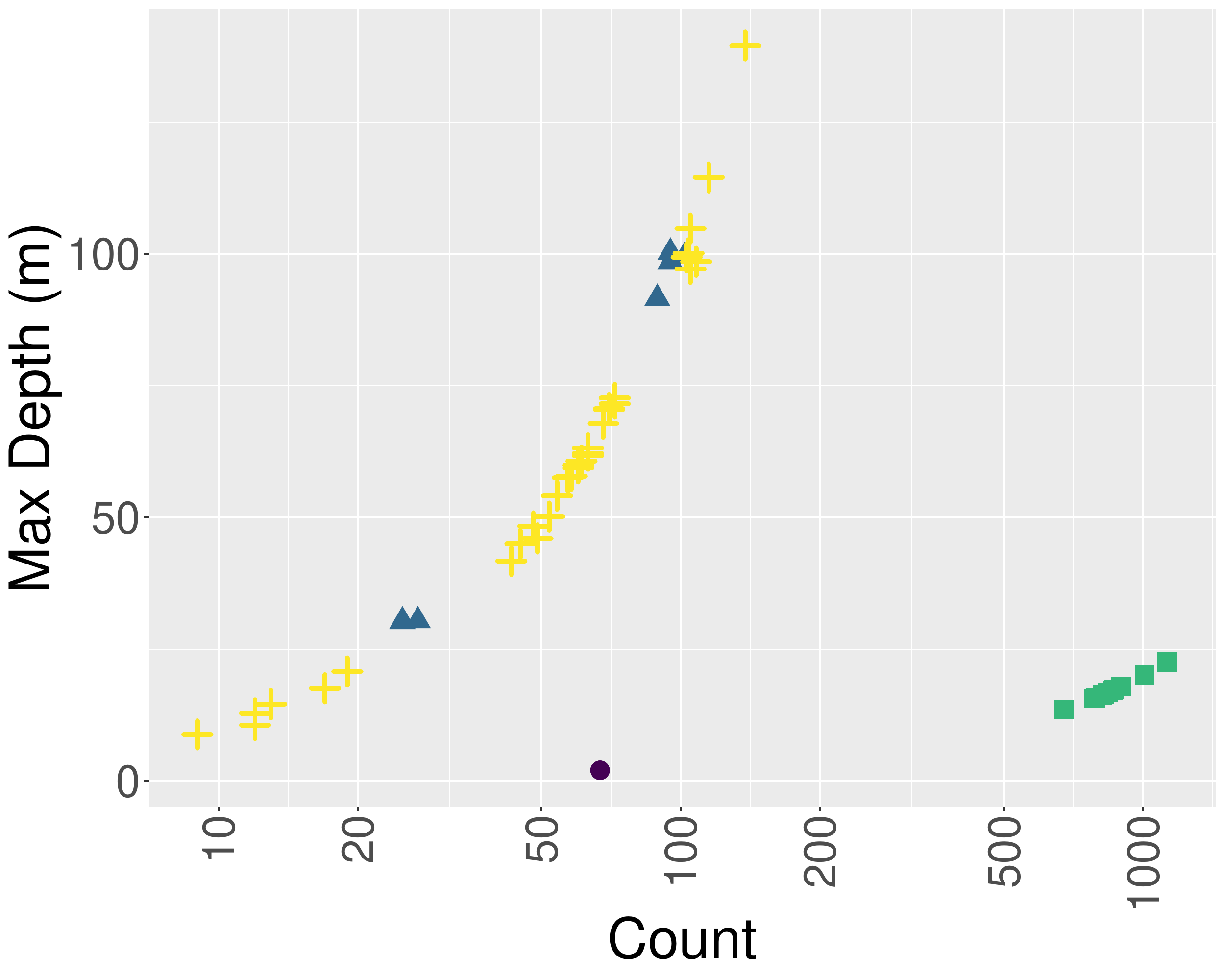}
\includegraphics[width=\textwidth]{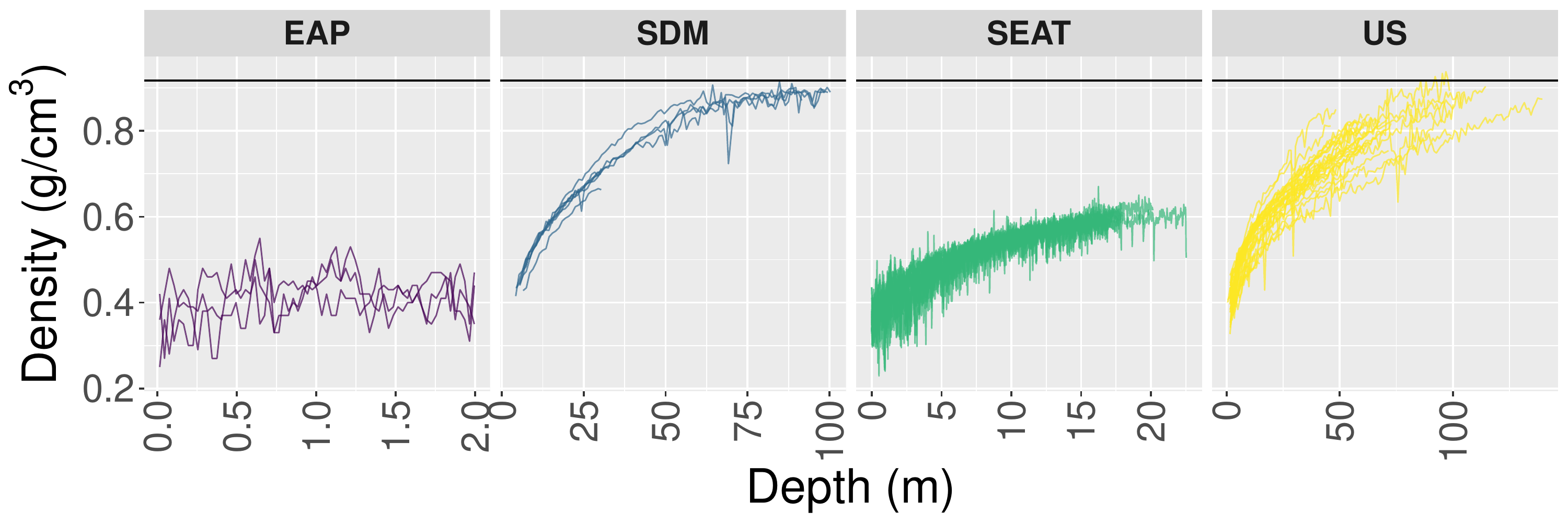}

\end{center}
        \vspace{-3mm}
\caption{(Top-Left)  Location of core sites with colors indicating campaign. The gray line shows flight lines with ice-penetrating radar measurements, and arrows labeled with letters A, B, and C, indicate sites where we illustrate snow density and SMB estimation. (Top-Right) Number of density measurements at each core plotted against core length (maximum depth) in m. (Bottom) Density measurements over depth by core, grouped by campaign. Note that the depth scales of these measurement types differ greatly. The black line indicates the density of solid ice.}\label{fig:locs}
        \vspace{-3mm}
\end{figure}

Here, we discuss the data characteristics and constraints that influence our modeling decisions. Our dataset consists of 57 snow/ice cores at $n_s = 56$ locations, each with many density measurements. We index density measurements $\rho(\bs_i,x)$ by the core $\bs_i$, where $\bs$ indicates the location and $i$ indexes replicattion at site $\bs$, and by depth $x$. We let $\mathcal{S}$ denote the collection of core sites. The locations of the snow cores $\bs_i \in \mathcal{S}$, the flight lines with ground penetrating radar, the number of measurements in each core $n_{\bs_i}$, the length of the core $x_{max,\bs_i}$, and the measured density $\rho(\bs_i,x)$ as a function of depth are plotted in Figure \ref{fig:locs}. Most cores have between 20 and 1,000 measurements, and, in total, the dataset contains $N = 14,844$ measurements.

These cores come from four field campaigns (indexed by $c(\bs_i)$), namely the East Antarctic Plateau \citep{albert_extreme_2004}, the Siple Dome project \citep{lamorey_waiscores}, the Satellite Era Accumulation Traverse \citep{burgener_observed_2013}, and the US portion of the International Trans-Antarctic Scientific Expedition \citep{mayewski_international_2005}.
In this paper, we refer to these different campaigns as EAP, SDM, SEAT, and US, respectively.
As these projects were distinct undertakings, precise methods and techniques of density measurement differ somewhat between them, although most are similar.

The traditional and most commonly used method involves measuring the mass and volume of core sections.
%Uncertainties in these estimates of density are approximately 5\% \citep{burgener_observed_2013}.
Density variability, however, relies not only on the mass and volume measurements themselves but also on the length of the core section used. 
Longer core sections smooth higher-frequency changes into a single bulk estimate, while increasingly small core sections better resolve short-term fluctuations.
Density measurements made in the field are necessarily on long sections (typically 1 meter), while measurements performed in a lab are often over a few centimeters.
%\cite{breton_design_2009} describes a relatively new technique to measure core density using gamma-ray attenuation with an unprecedented resolution of 3.3 mm.
The accuracy and precision of density measurements, therefore, depend on both the method and resolution of the measurements.
Given the data attributes discussed, we consider combinations on variance models that account for (1) the length of the core that is used to obtain that measurement and (2) the field campaign that obtained the density measurement. 

\begin{figure}[ht]
    \centering
        \includegraphics[width=\textwidth]{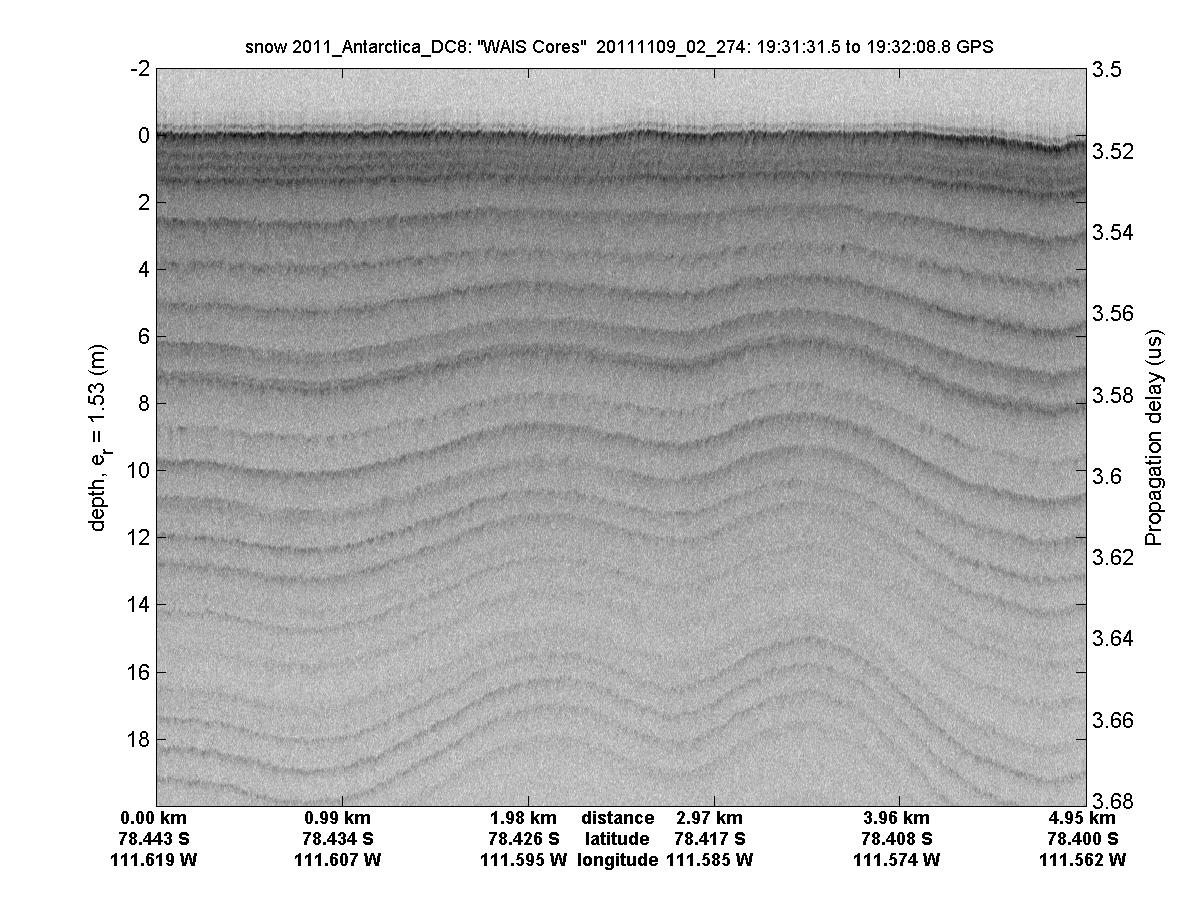}
    \caption{Example of an ice-penetrating radar image. The visible layering represents seasonal variations in snow properties, causing reflections of radar pulses back to the transceiver. These annual layers, combined with snow depth-density estimates, are used to estimate annual SMB. Image from \citep{Paden_2014}.}\label{fig:echogram}
\end{figure}

Ice-penetrating radar permits non-destructive imaging of the snow sub-surface.
Seasonal variations (in, e.g., air temperature, dust deposition, compaction rates, etc.) result in differences in the electromagnetic properties of snow deposited in summer compared to that in winter \citep{alley_summertime_1990}.
These contrasts act as reflection horizons for electromagnetic pulses sent from ground-based or airborne active radar systems \citep{fujita_nature_1999}.
The combined reflected radar signals from repeated pulses of a moving radar system can, therefore, image the subsurface annual layering in Antarctic snow (see Figure \ref{fig:echogram}).

\section{Monotone Integrated Spatial Process}\label{sec:models}

In this section, we present the class of monotone integrated spatial processes (MISP) for estimating monotone snow density curves. We define $z(\bs,x) > 0$ to be a positive space-depth analog to $z(x)$ in \eqref{eq:sol_diff_eq}. Using the integral of $z(\bs,x)$, with respect to depth, we model the mean function of snow density $\mu(\bs,x)$ as
\begin{equation}\label{eq:MISP}
   \log\left( \frac{\mu(\bs,x)}{ \rho_I - \mu(\bs,x)} \right) =w(\bs,x) = \alpha(\bs) + \int_0^x z(\bs,t) dt.
\end{equation}
This constrains the mean function $\mu(\bs,x) \in (0,\rho_I)$ to be monotone increasing. To explore its properties here and in the Supplemental Material, we define $w(\bs,x)$ to be the untransformed MISP. We clarify, however, that we only use $\mu(\bs,x)$, a transformation of non-Gaussian processes, as the mean snow density.

As a brief aside, we mention a few of the mathematical properties of the MISP. Because $z(\bs,x) > 0$, for any $x' > x$, we know $w(\bs,x') > w(\bs,x)$. That is, $w(\bs,x)$ is monotone increasing as a function of $x$ for any location $\bs$. By standard real analysis, because $w(\bs,x)$ is a monotone function over any interval $(0,T)$, it is differentiable almost everywhere on that interval by Lebesgue's theorem for the differentiability of monotone functions. 
In addition to being differentiable almost everywhere, the covariance of $w(\bs,x)$ is nonseparable and depth non-stationary, and there are explicit relationship between the mean and covariance \citep{hefley2017basis} (See the Online Supplement). Thus, this model provides a very flexible class of monotone functions. %Thus, $w(\bs,x)$ is a \emph{smooth} random effect, regardless of the specification of $z(\bs,x)$.

To construct a positive space-depth process $z(\bs,x)$, we use kernel convolutions \citep{higdon1998,higdon2002}. The integrated space-depth function is
\begin{equation}\label{eq:kern}
\begin{aligned}
\int_0^x z(\bs,t)\,  dt &=  \int_0^x \int_\mathbb{R}  k(t - u)  z^*(\bs,u) \,du \, dt, \\
&=  \int_\mathbb{R} \int_0^x k(t - u)  z^*(\bs,u)\, dt  \,du,  \\
&= \int_\mathbb{R} \left(K(x - u)  - K(0 - u)\right) z^*(\bs,u)  \,du ,
\end{aligned}
\end{equation}
where $k(\cdot)$ is the smoothing kernel (a scaled PDF) with corresponding scaled CDF $K(\cdot)$ and $z^*(\bs,u)$ is a spatial LGP for every depth $u$. Thus, our proposed method is a kernel convolution model using differences of CDFs as kernels. For computational feasibility with $ N > 10^4$ and because the mean density function appears relatively smooth, we reduce the depth dimensionality of the space-depth process using process convolutions with a selection of depth-ordered knots $x_{min} = 0 \leq \xi_1 \leq... \leq \xi_J \leq x_{max} = \max_{\bs_i} x_{max,\bs_i}$. However, we do not reduce dimensionalilty over space. In this framework, we can express the space-depth random function $w(\bs,x)$ as a linear model,
\begin{equation}\label{eq:MISP_approx}
\begin{aligned}
w(\bs,x) &= \alpha(\bs) + \sum^J_{j = 1} K_j(x) z_j^*(\bs) \\ &= \alpha(\bs) + \bK(x)^T \bz^*(\bs)
\end{aligned}
\end{equation}
where $K_j(x) = \int_0^x k_j(u - \xi_j) \, du$, and $k_j(\cdot)$ and $z_j^*(\bs)$ are kernels and independent spatial log-Gaussian processes associated with each knot $\xi_j$. The vectors $\bK(x)$ and $\bz^*(\bs)$ contain $J$ elements of integrated kernels $K_j(x)$ and log-Gaussian process elements $z_j^*(\bs)$ for $j = 1,..,J$. Importantly, the kernels $k_j(\cdot)$ need not be symmetric, and there are potential computational benefits to using truncated kernels as this induces sparseness in $\bK(x)$.

To complete this model, we must specify several components: the number of knots $J$, the spatially-varying intercept $\alpha(\bs)$, the smoothing kernels $k_j(\cdot) > 0$, and the spatial log-Gaussian processes $z_j^*(\bs)$, $j = 1,...,J$. The properties of this space-depth model are determined by these selections. We consider specifying $k_j(\cdot)$ using Normal, $t$, and Asymmetric Laplace probability distributions with full support ($\mathbb{R}$), as well as M-spline bases. M-spline bases are in fact probability densities as they are scaled to integrate to one \citep{ramsay1988}, and, when integrated, M-splines bases yield I-spline bases common to monotone regression. Thus, monotone I-splines models are a special case of \eqref{eq:MISP_approx} and induce sparseness in $\bK(x)$. M-splines bases naturally induce sparseness, but using truncated probability distributions can provide the same desirable sparseness in ``big-data'' settings. We discuss the M-spline basis function in more detail in the Supplementary Material.
We make model selections based on out-of-sample predictive performance (See section \ref{sec:modcomp}).

\section{Methods and Models}\label{sec:mod}

\subsection{Model Selection}\label{sec:modcomp}

To compare different models, we carry out 19-fold cross-validation for each model considered because our data come from 57 snow cores, and holding out three cores each model fitting is a convenient choice. Our modeling goal is estimating the entire snow density function at locations without drilled snow cores. In our comparison, therefore, we weight predictive performance measures by the length of the core that each hold-out measurement represents. Our model comparison measures, therefore, correspond to approximated integrated error measures common to density or function estimation \citep[see, e.g.,][]{fryer1976,marron1992}. 

We let all model parameters be $\btheta$ and use Markov chain Monte Carlo (MCMC) to obtain $M$ posterior samples from the posterior distribution of $\btheta$. For each posterior sample $\btheta^{(m)}$, $m = 1,...,M$, from our Markov chain Monte Carlo model fitting, we generate a corresponding prediction $\rho^{(m)}(\bs_i,x)$ for each hold-out observation $\rho(\bs_i,x)$ from the posterior predictive distribution. We propose several criteria for comparing predictions to hold-out data: predictive squared and absolute error, as well as a strictly proper scoring rule \citep{gneiting2007}, the continuous ranked probability score (CRPS) \citep[see][for early discussion on CRPS]{brown1974,matheson1976}.  
We estimate CRPS using the empirical CDF of posterior predictions $\rho^{(m)}(\bs_i,x)$ \citep{kruger2016predictive}, which we denote $\text{CRPS}(\hat{F}(\bs_i,x),\rho(\bs_i,x))$,
\begin{equation}
\frac{1}{M} \sum_{j=1}^M \mid \rho^{(j)}(\bs_i,x) - \rho(\bs_i,x) \mid  - \frac{1}{2M^2} \sum_{m=1}^M \sum_{m'=1}^M \mid  \rho^{(m)}(\bs_i,x) - \rho^{(m')}(\bs_i,x) \mid,
\end{equation}
where $\rho^{(m)}(\bs_i,x)$ represents the $m$th sample from the posterior predictive distribution for core $\bs_i$ and depth $x$. Unlike squared or absolute error which only use the posterior mean, CRPS compares the entire posterior predictive distribution to hold-out values and rewards predictive distributions concentrated on the correct value.

To define integrated squared and absolute errors, we use the maximum depth $x_{max,\bs_i}$ of the core $\bs_i$ and number of measurements $n_{\bs_i}$ at $\bs_i$. We define integrated squared error (ISE) as
\begin{equation}
\sum_{\bs \in \mathcal{S} } \sum_i \frac{x_{max,\bs_i}}{n_{\bs_i}} \sum_{x} \left( \frac{1}{M} \sum^M_{m=1} \rho^{(m)}(\bs_i,x) - \rho(\bs_i,x)  \right)^2.
\end{equation}
Similarly, we defined integrated absolute error (IAE)
\begin{equation}
\sum_{\bs \in \mathcal{S} } \sum_i \frac{x_{max,\bs_i}}{n_{\bs_i}} \sum_{x} \left| \frac{1}{M} \sum^M_{m=1} \rho^{(m)}(\bs_i,x) - \rho(\bs_i,x) \right|.
\end{equation}
We do not define an integrated (or weighted) CRPS because CRPS naturally incorporates model uncertainty (including weighting) probabilistically by accounting for the entire posterior predictive distribution.

We now present a summary of our model comparison, deferring a complete outline of the results to the Supplementary Material. The final model uses piecewise constant M-spline kernels $k_j(\cdot)$, corresponding to a linear I-spline, with five interior knots at 5, 15, 30, 45, and 75 meters. To write this model as \eqref{eq:MISP_approx}, we must include an additional knot at 0 meters. The knot locations were chosen through model selection among a variety of knot selections. These depth knots capture expected changes in densification patterns due to the interplay of particle rearrangement and plasticity changes at critical densities \citep{herron_firn_1980,horhold_densification_2011}. The spatially-varying intercept $\alpha(\bs)$ and the log-Gaussian processes $z_j^*(\bs)$ are specified independently with exponential covariance using the great-circle distance, denoted $d(\bs,\bs')$, with shared spatial decay parameter $\phi$ and unique scale parameters $\sigma^2_j$. The Gaussian process intercept and log-Gaussian process coefficients are centered on a unique scalar means. The mean function is nested within a truncated-Normal distribution with campaign-specific variance, scaled by the length of the core each measurement represents. The model is described in detail in Section \ref{sec:hierarchical} with prior distributions discussed in Section \ref{sec:prior}.

\subsection{Hierarchical Model}\label{sec:hierarchical}

Here, we present the model for snow density with the best out-of-sample predictive performance. We constrain the mean function to lie between 0 g/cm$^3$ and the density of ice $\rho_I$ using a generalized logistic function with $\rho_I$ as a maximum. Our hierarchical model for snow density is
\begin{equation}\label{eq:final_mod}
\begin{aligned}
\rho(\bs_i,x) &\sim \mathcal{TN}\left( \mu(\bs,x),\tau^2_{c(\bs_i)}\frac{n_{\bs_i}}{x_{max,\bs_i}},0,\infty \right),  &\rho(\bs_i,x) > 0 \\
 \log \left( \frac{\mu(\bs,x)}{\rho_I - \mu(\bs,x)} \right) &= \alpha(\bs) + \bK(x)^T\bz^*(\bs), &\mu(\bs_i,x) \in (0,\rho_I] \\
 \alpha(\bs) &\sim \text{GP}\left(\gamma_0, \sigma^2_0 \exp\left( -\phi d(\bs,\bs') \right)\right) &\alpha(\bs)\in \mathbb{R} \\
z_j^*(\bs) &\overset{ind}{\sim} \text{LGP}\left(\gamma_j, \sigma^2_j \exp\left( -\phi d(\bs,\bs') \right)\right) &z_j^*(\bs) > 0
 \end{aligned}
\end{equation}
where $\mathcal{TN}(\cdot,\cdot,0,\infty)$ is a Gaussian error model truncated below by 0 using core-specific weights constructed with the maximum depth $x_{max,\bs_i}$ and the number of density measurements $n_{\bs_i}$, scaled by a campaign-specific scale parameter $\tau^2_{c(\bs_i)}$. While the truncation is necessary to provide the proper support for $\rho(\bs_i,x) >0$, in practice, the truncation does very little because observations are not close to 0 g/cm$^3$ given that estimated variances are small. Therefore, we refer to $\mu(\bs,x)$ as a mean function rather than a location function. The GPs for $\alpha(\bs)$ and $\log(z_j(\bs))$ are independent with shared decay parameter $\phi$, unique scale parameters $\sigma^2_0$,...,$\sigma^2_6$, and location parameters $\gamma_0$,...,$\gamma_6$. Importantly, because each $z_j^*(\bs)$ is a log-Gaussian process, they each have a multiplicative rather than additive errors. Here, the mean of the model is a scaled inverse-logit tranformation of a MISP using M-spline kernels with five interior knots (six knots in terms of \eqref{eq:MISP_approx} discussed in Sections \ref{sec:models} and \ref{sec:modcomp}). The mean function $\mu(\bs,x)$ is not indexed by $i$ because it does not depend on the core; however, the variance of the model is dependent on the core $\bs_i$ and the campaign $c(\bs_i)$ that analyzed the core.

\subsection{Prior Distributions, Model Fitting, and Interpolation}\label{sec:prior}

To complete the model, we specify prior distributions for all model parameters. In this setting, standard mean-zero prior distributions for $\gamma_{0},...,\gamma_{6}$ would suggest a model with high surface density and rapid snow densification. Our goal in selecting a prior distribution was choosing a model that would produce very flexible snow density estimates. Here, we assume the following prior distributions:
\begin{equation}
    \begin{aligned}[c]
    \gamma_0 &\sim \mathcal{N}(-0.5,1),  &\hspace{1mm} \\
            \sigma^2_0 &\sim \mathcal{IG}(10,3),  &\hspace{1mm}  \\
    \phi &\sim \text{Uniform}(10^{-5},10^{-1}),  &\hspace{1mm} 
    \end{aligned}
    \hspace{2mm}
        \begin{aligned}[c]
            \gamma_j &\sim \mathcal{N}(-1.5,1),  &\text{ for } j = 1,...,6 \\
    \sigma^2_j &\sim \mathcal{IG}(4,3),  &\text{ for } j = 1,...,6  \\
        \tau^2_{c(\bs_i)} &\sim \text{Gamma}(1,100),  &\text{ for all } c(\bs_i) &\hspace{1mm} 
    \end{aligned}
\end{equation}
where $\mathcal{N}$, $\mathcal{IG}(.,.)$, and $\text{Gamma}$ are Normal, Inverse-Gamma, and Gamma Distributions, respectively. Here, we use the parameterization of $\mathcal{IG}(a,b)$ indicating a mean $b/(a-1)$ and $\text{Gamma}(a,b)$ that has expectation $a/b$.

\begin{figure}[ht]
        \vspace{-3mm}

    \centering
        \includegraphics[width = 0.4\textwidth]{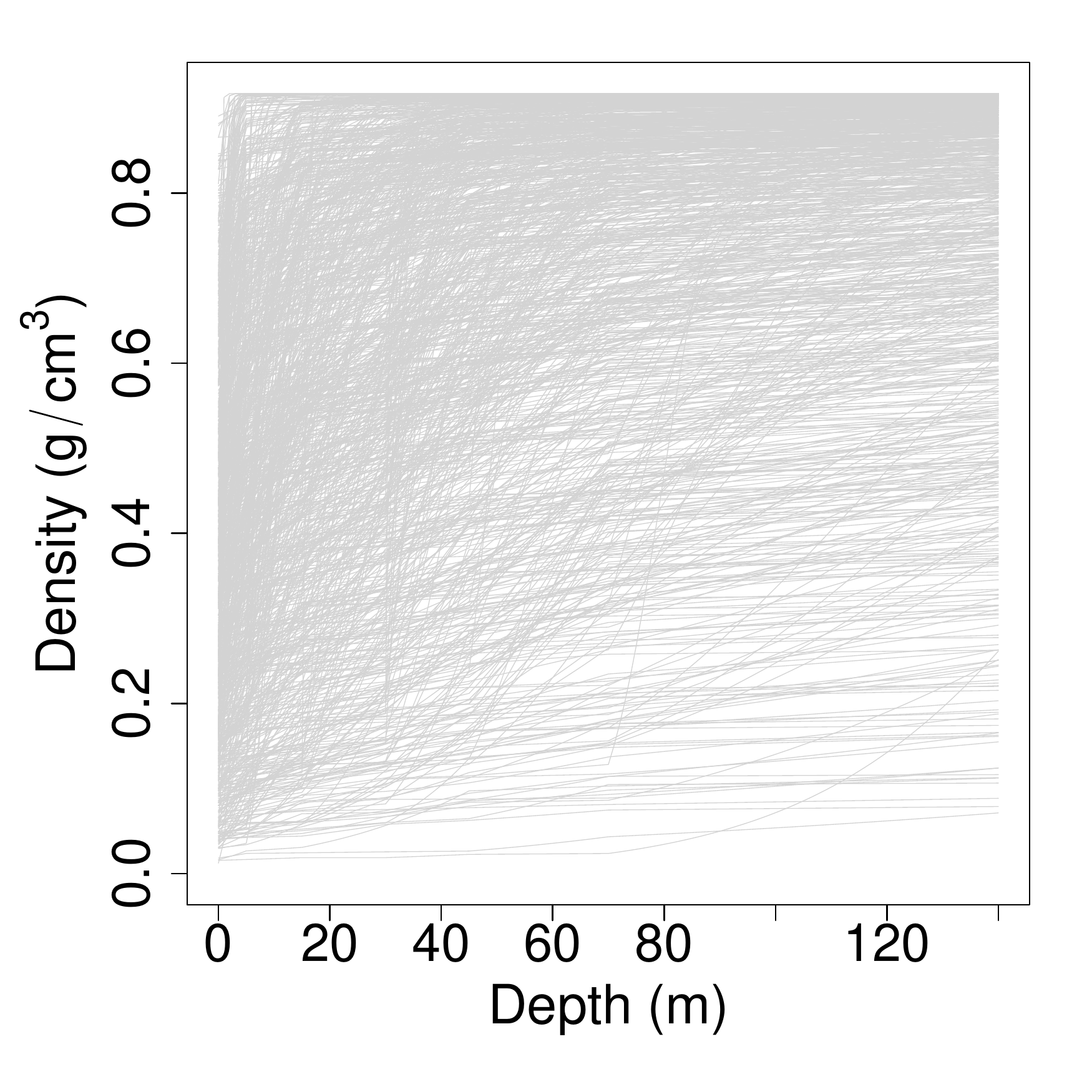}
        \includegraphics[width = 0.4\textwidth]{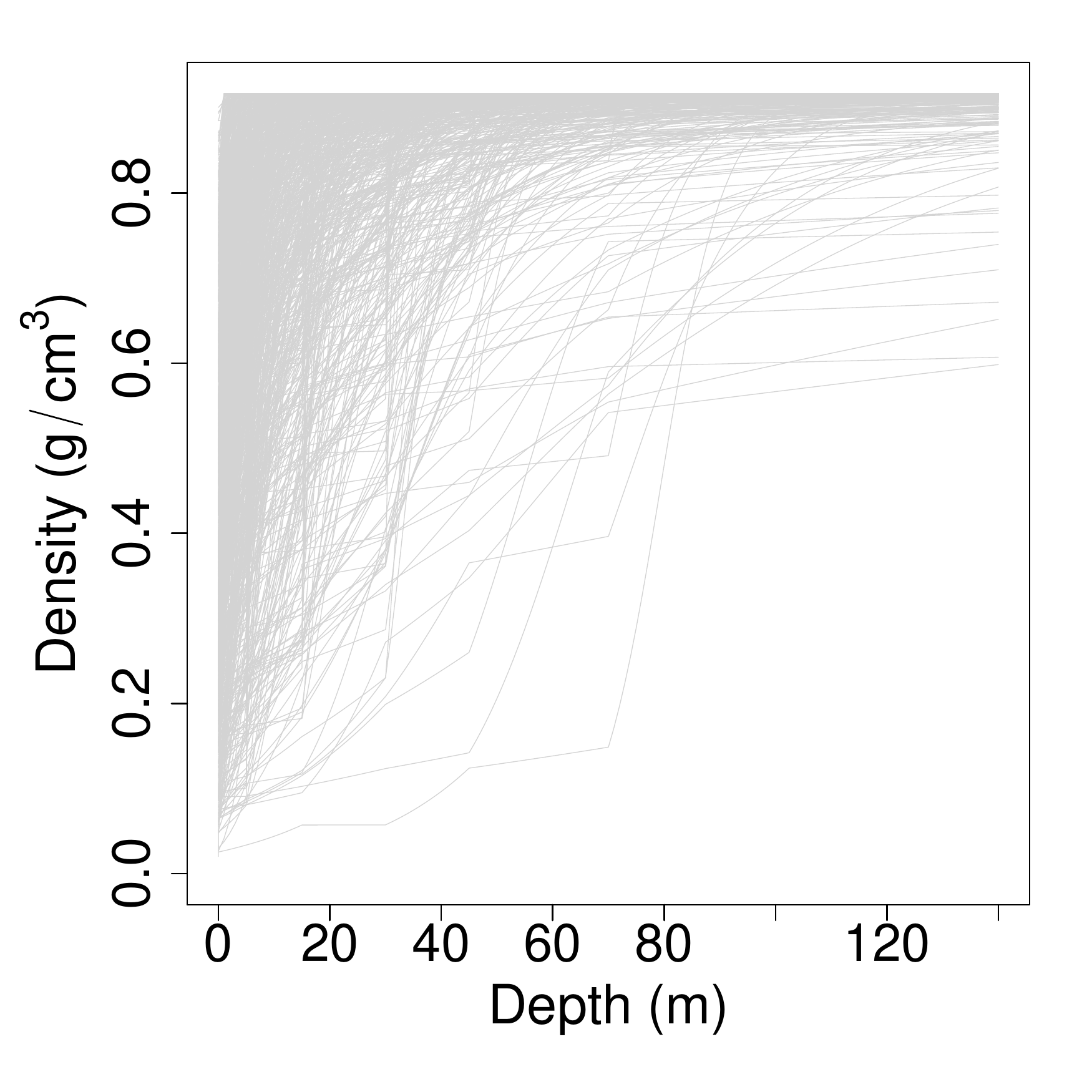}
        \vspace{-3mm}
    \caption{Prior predicted mean snow density curves under (Left) our proposed model and (Right) zero-mean prior distributions.}    \label{fig:priorpred}
        \vspace{-3mm}

\end{figure}

Although these prior distributions are informative and non-standard, we select them so that the mean prior surface density is between 0.35 - 0.4 g/cm$^3$ and so that simulations from the prior distribution yield plausible and flexible snow density curves. In Figure \ref{fig:priorpred}, we plot 1,000 realizations of the mean snow density simulated from our prior distribution and under a more standard, mean-zero model. Note that the zero-mean model generates informative and unrealistic density curves that put prior weight on near-ice density at shallow depths. Thus, our seemingly more informative prior is less informative and more realistic in the data space, while allowing great flexibility. We also highlight that there are several large jumps in the samples from the non-informative specification because the mean-zero normal random variables, when exponentiated, can be quite large and induce very rapid changes in the mean function. These changes in the function are particularly visible at knot locations.
Lastly, we choose the prior distribution for $\phi$, the common decay parameter for $\alpha(\bs)$ and $\log(z_j(\bs))$, to be uniform between $10^{-5}$ and $10^{-1}$ to allow a wide range of possible values. 

We sample from the posterior distribution, $\pi( \btheta | \brho)$, where $\brho$ denotes all density measurements, using Hamiltonian Monte Carlo (HMC), implemented in Stan \citep{carpenter2017}. While several model parameters can be sampled in closed form using a Gibbs Sampler, the spatial random effects cannot, and using HMC improved the mixing of spatial random effects. Letting $\btheta$ be all model parameters, this model fitting approach yields $M$ samples from the posterior distribution.

We use each of the posterior samples ($\btheta^{(1)},...,\btheta^{(M)}$) to estimate snow density at hold-out locations and depths sampling from the posterior predictive distribution,
% \begin{equation}\label{eq:post_pred}
%f(\rho(\bs,x) | \brho) = \int f ( \rho(\bs,x) |\btheta) \, \pi( \btheta | \brho) \, \textrm{d}\btheta,
%\end{equation}
via composition sampling \citep{tanner1996}. That is, for every posterior sample $\btheta^{(m)}$, we simulate from the data model \eqref{eq:final_mod} to get posterior predictions for every hold-out $\rho(\bs_i,x)$. When estimating snow densities at unobserved locations, however, we estimate $\mu(\bs,x)$ rather than a noisy version of the mean. In conjunction with ice-penetrating radar measurements, estimated snow density curves can then be used to estimate a history of surface mass balance at that location using the methods discussed in Section \ref{sec:firn}, as we demonstrate in Section \ref{sec:res}.

%When comparing models using criteria from Section \ref{sec:modcomp}, we simulate posterior predictions from \eqref{eq:post_pred} by generating realizations of the mean function $\mu^{(m)}(\bs,x)$ for the hold-out core $\bs_i$ by sampling $\alpha(\bs)$ and $\log(z_j(\bs))$from their posterior conditional normal distribution, conditioning, respectively, on all other $\alpha(\bs)$ and $\log(z_j(\bs))$ at $\bs \in \mathcal{S}$. Then, we sample from the corresponding truncated normal in \eqref{eq:final_mod} to obtain posterior predictive samples. 

\section{Results}\label{sec:res}

Using Stan, we run our MCMC sampler for 55,000 iterations. We discard the first 5,000 iterations, yielding 50,000 posterior samples on which we base our posterior inferences. 
%After convergence, thinning increases posterior variance \citep{geyer1992,maceachern1994}; however, in this setting, we thin to decrease memory requirements while carrying out posterior inference and because we do not expect too much information loss since spatially-correlated parameters $\alpha(\bs)$ and $z_j(\bs)$ mix slowly. 
In Supplemental Material, we demonstrate that the MCMC is well-behaved. In total, this model fitting takes approximately 24 hours using one Intel(R) Xeon(R) Gold 6142 CPU @ 2.60GHz processor.

\subsection{Posterior Summaries}

\begin{figure}[ht]
    \centering
    \includegraphics[width = 0.9\textwidth]{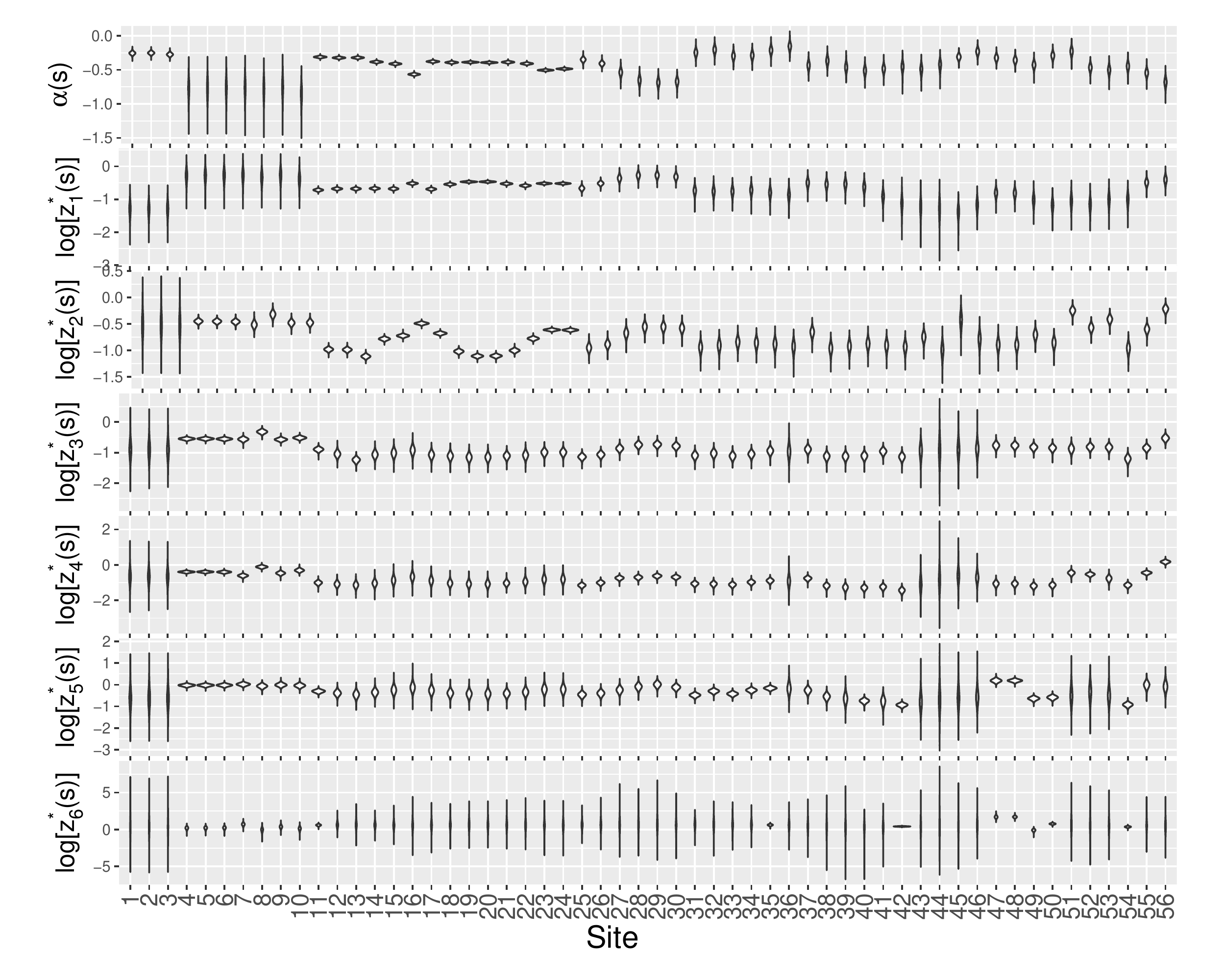}
            \vspace{-3mm}
    \caption{Violin plots of posterior samples for all spatially-distributed random effects. } \label{fig:random_effects}
        \vspace{-3mm}
\end{figure}

Based on 50,000 posterior samples, we provide violin plots for the posterior distributions for spatial terms, $\alpha(\bs)$ and $\log(z_j^*(\bs))$, in Figure \ref{fig:random_effects}. We defer other posterior summaries to the Supplemental Material but note that the campaign-specific variances differ significantly, suggesting that these campaigns contributed different levels of noise to the data. 

For some sites, the posterior distributions are more diffuse, particularly for coefficients ($z_j^*(\bs)$) that correspond to I-spline bases at greater depths. This generally happens when a snow core is shallow and does not extend past one or more of the interior knots. For example, the EAP cores are 2-meter snow pits. For such cores, as discussed in Section \ref{sec:models}, estimation of $\bz^*(\bs)$ relies on information shared from nearby cores that have deeper observations. In addition, when the estimated snow density $\mu(\bs,x)$ is close to $\rho_I$, even large changes in $z_j^*(\bs)$ may have little effect on $\mu(\bs,x)$. For both of these reasons, estimates of $z_j^*(\bs)$ are generally more variable for larger $j$ (See Figure \ref{fig:random_effects}).

The SDM cores, labeled 4-10, have lower and more uncertain estimated intercepts $\alpha(\bs)$ than other sites, meaning that the estimated surface density is lower relative to other sites. These cores are geographically, and thus climatically, isolated. For this reason, it is unsurprising that appear different than other cores, but these estimates are not incompatible with the estimates of other cores, cores 28-30, for example.

\subsection{Interpolation and Extension of Snow Density Curves}

In this section, we show two utilities of this model: (1) extending snow density estimates beyond the depth of the original core and (2) estimating snow density curves at locations where snow cores have not been drilled. Both tasks are scientifically important as the first task aids in studying a longer history of the Antarctic ice sheets, while the second task allows us to estimate water accumulation in locations where we have not drilled snow cores. 

\begin{figure}[ht]
                        \vspace{-3mm}

    \centering
    \includegraphics[width = 0.32 \textwidth]{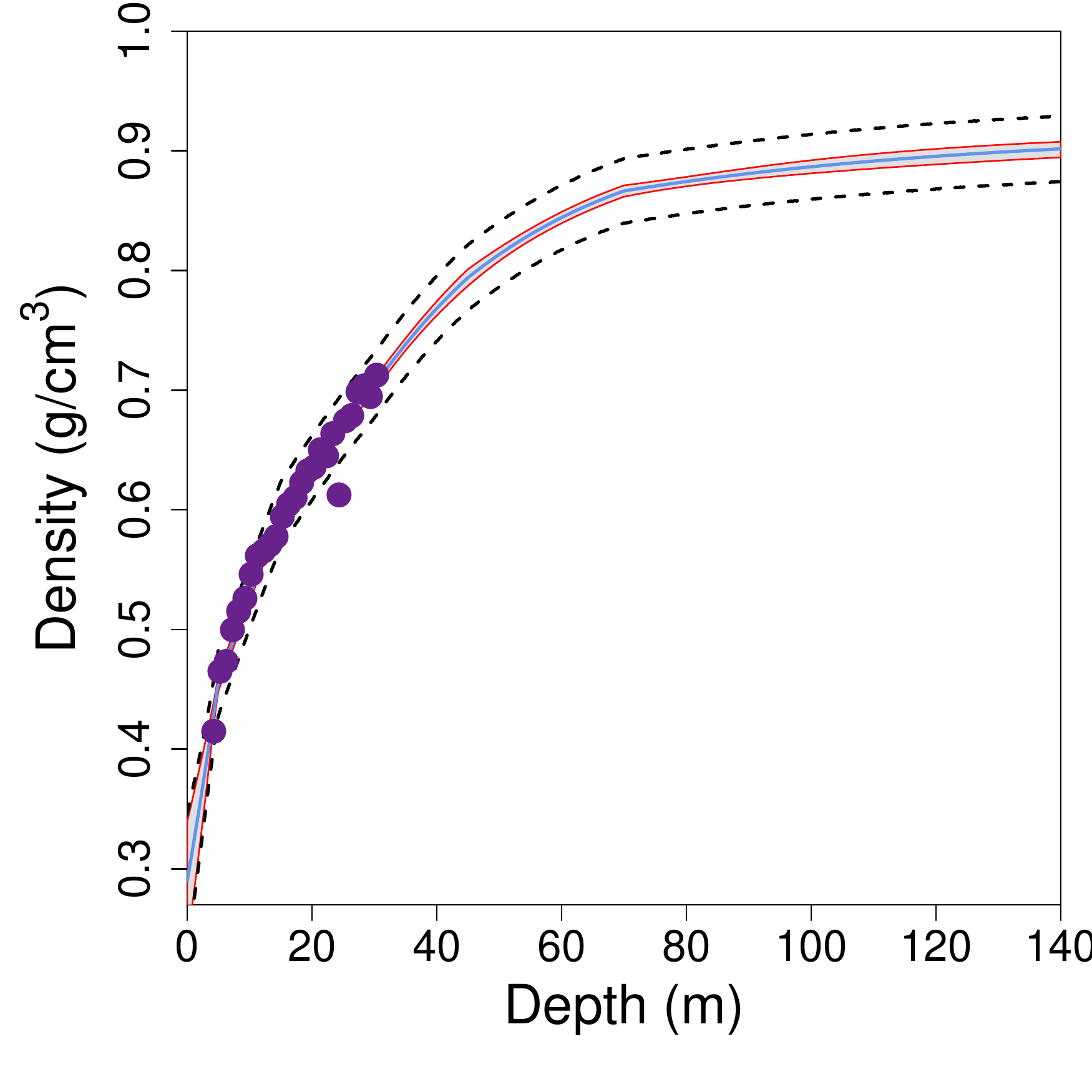}
        \includegraphics[width = 0.32 \textwidth]{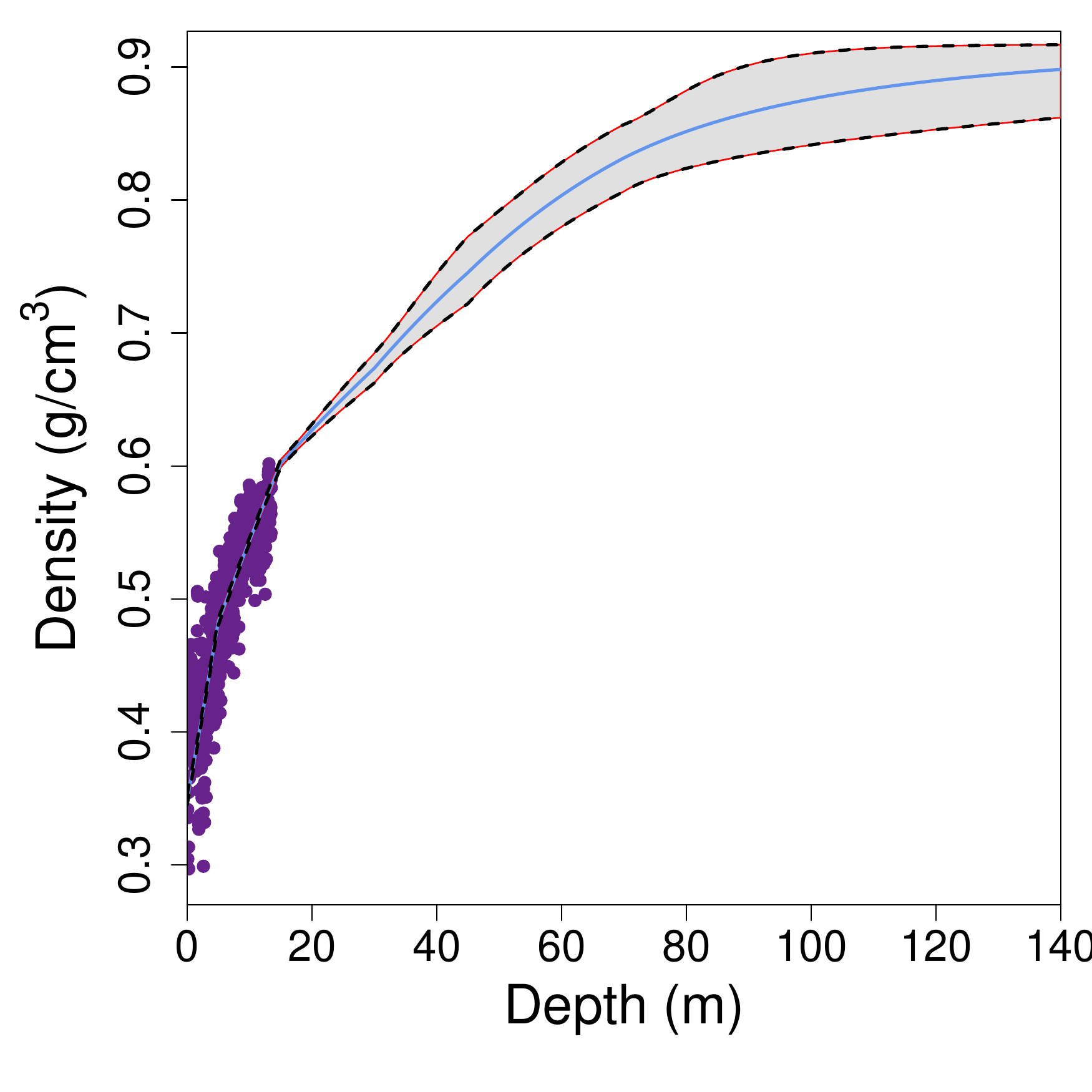}
        \includegraphics[width = 0.32 \textwidth]{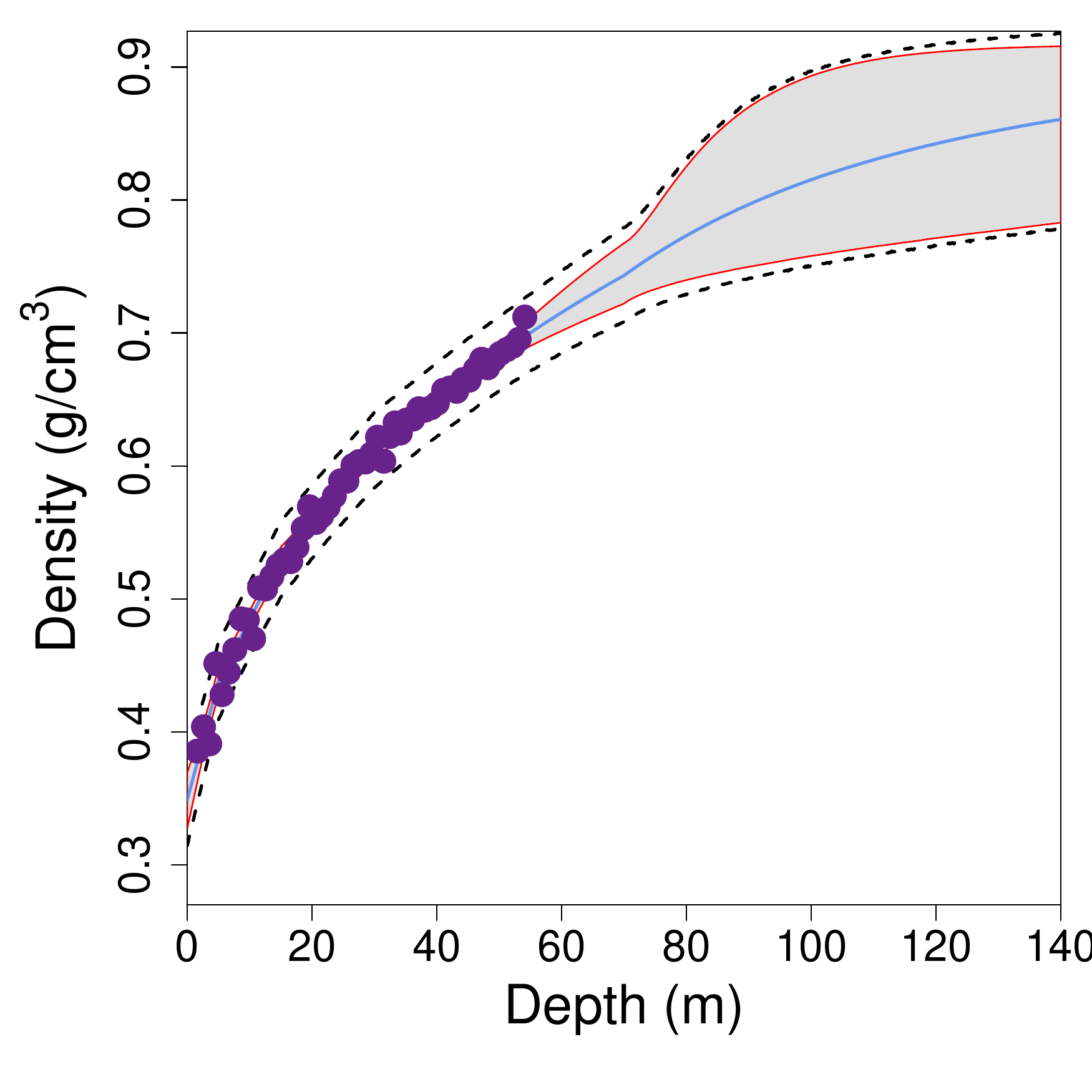}
                    \vspace{-3mm}

            \caption{Estimated snow density down to 140 m for (Left) Core 6, (Center) Core 24, and (Right) Core 42. The mean function is plotted in blue, 95\% credible intervals on the mean as a polygon with red boundaries, 95\% posterior prediction intervals in black dashed lines (using the correct campaign-specific scale parameter and assuming the same data spacing present in each core), and density measurements plotted as purple points.}    \label{fig:extend}

                        \vspace{-3mm}
\end{figure}

For the first task, we use cores 6, 24, and 42 (See Figure \ref{fig:locs}) as an illustration. Each core represents a unique scenario. Core 6 comes from a very tight cluster of deep snow cores from the SDM project drilled near the coast with neighboring cores extending to nearly 100 meters that aid in precise estimation of snow densities at deeper depths. In coastal areas, densification generally occurs more rapidly due to slightly higher temperatures; therefore, there is a smaller difference between the density of ice and observed densities, making density estimates more precise. Core 24 is a 13.5 meter SEAT core in a data-rich area; however, in this region, there are not many deep cores. Core 42 is a US snow core that is slightly longer than 50 meters but is not in a relatively data-rich area. For each core, we estimate the mean snow density $\mu(\bs,x)$ down to 140 meters (See Figure \ref{fig:extend}).

Because core 6 is near many deep cores, the estimated density curve is very precise even though it only extends to 30 meters. The estimated snow density estimates for core 24 are very precise to about 40 m but become less certain at greater depths because neighboring cores lend less information at those depths. Compared to cores 6 and 24, core 42 lies in a data-poorer region with only four other cores within a 500 km. Thus, our density estimates are much less certain beyond the range of measured density.

\begin{figure}[ht]
                        \vspace{-3mm}

    \centering
    \includegraphics[width = 0.32 \textwidth]{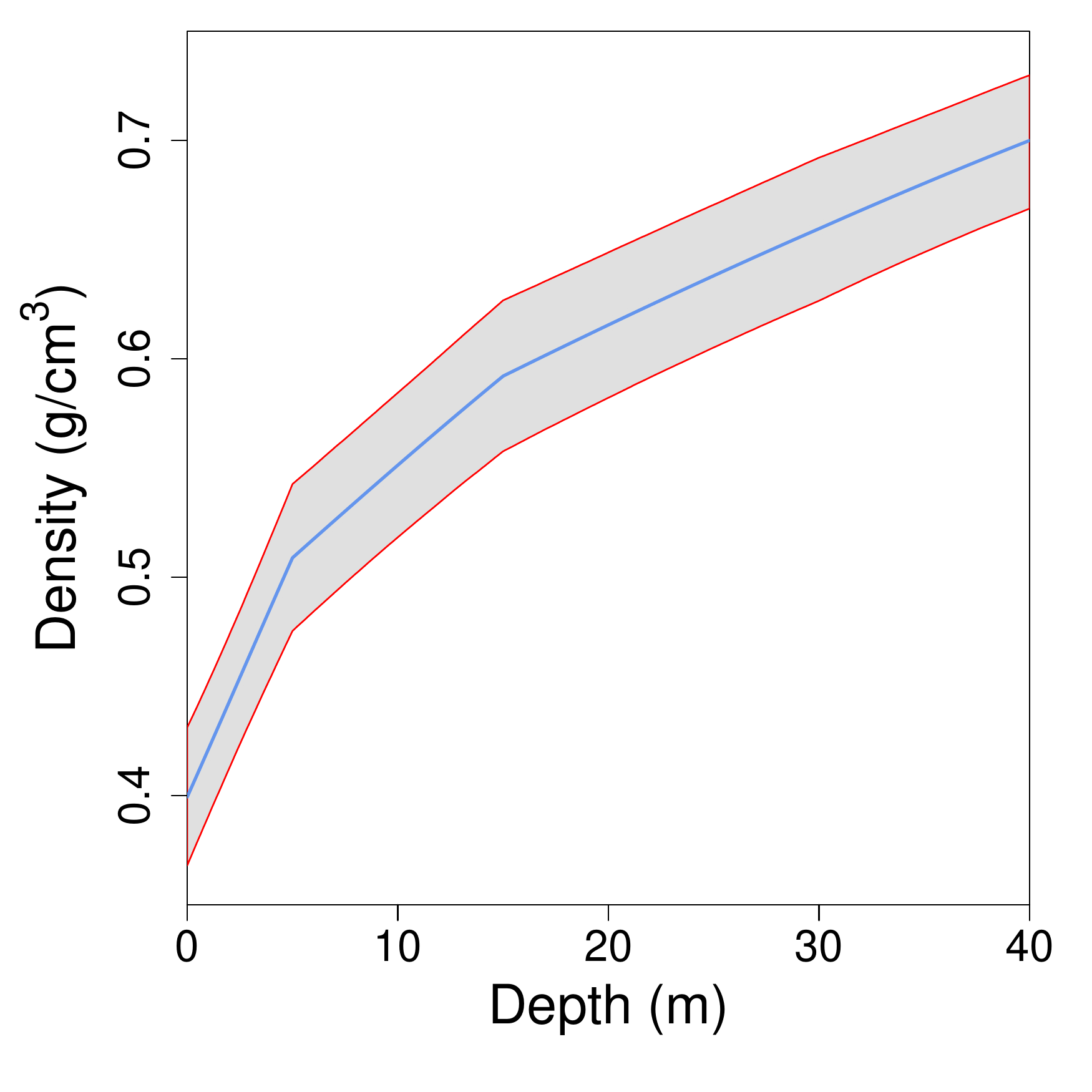}
        \includegraphics[width = 0.32 \textwidth]{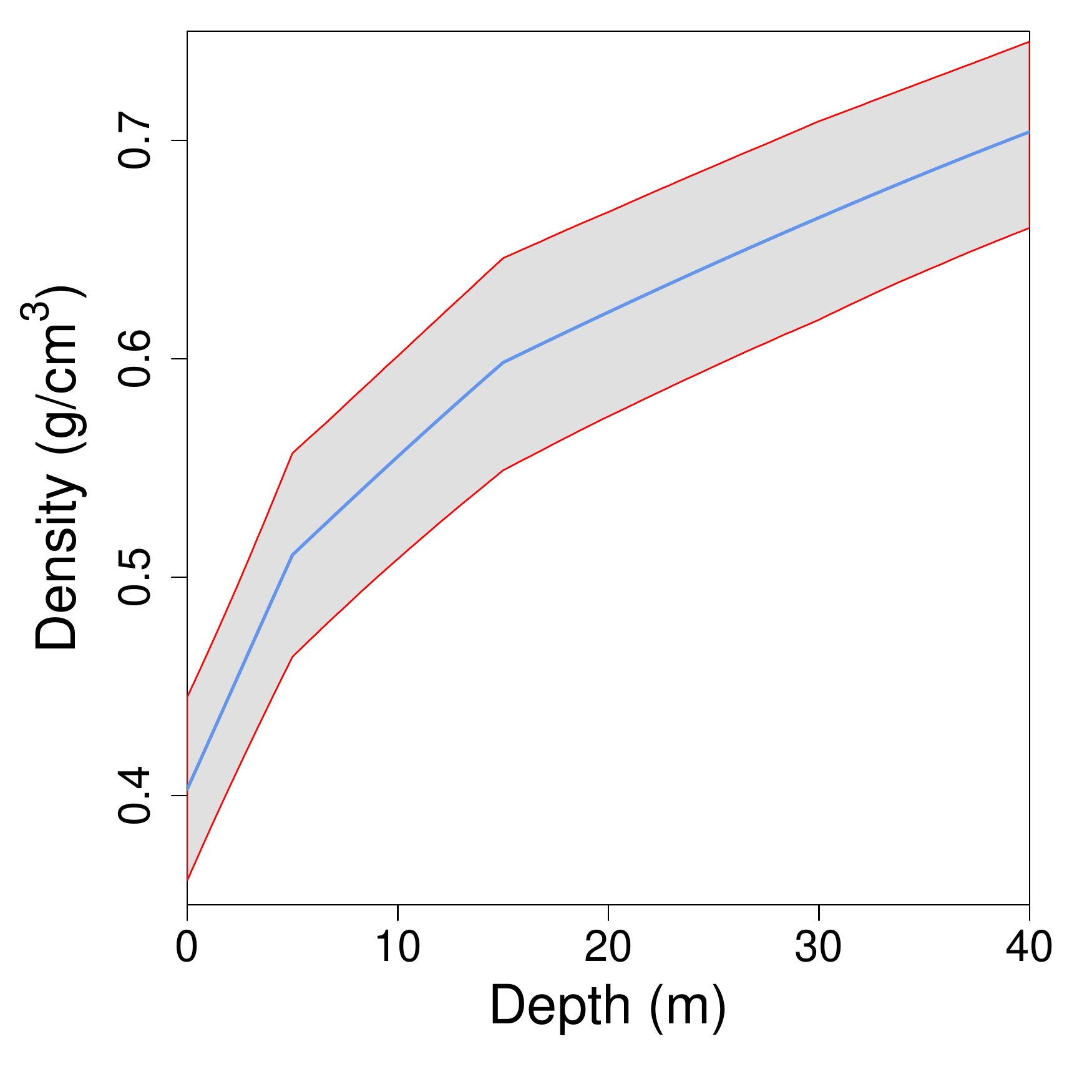}
             \includegraphics[width = 0.32 \textwidth]{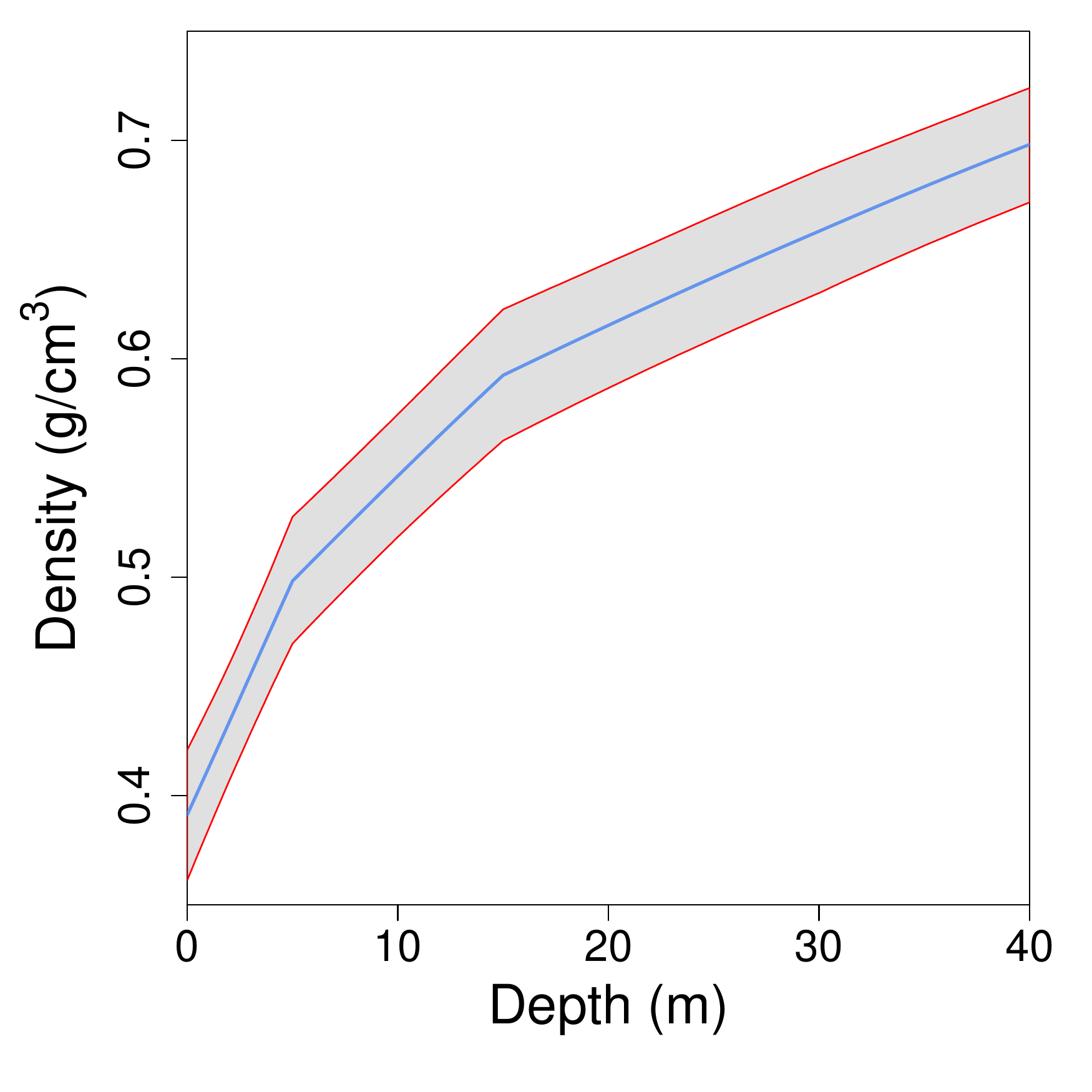}
             
             \vspace{-3mm}
             
    \includegraphics[width = 0.32 \textwidth]{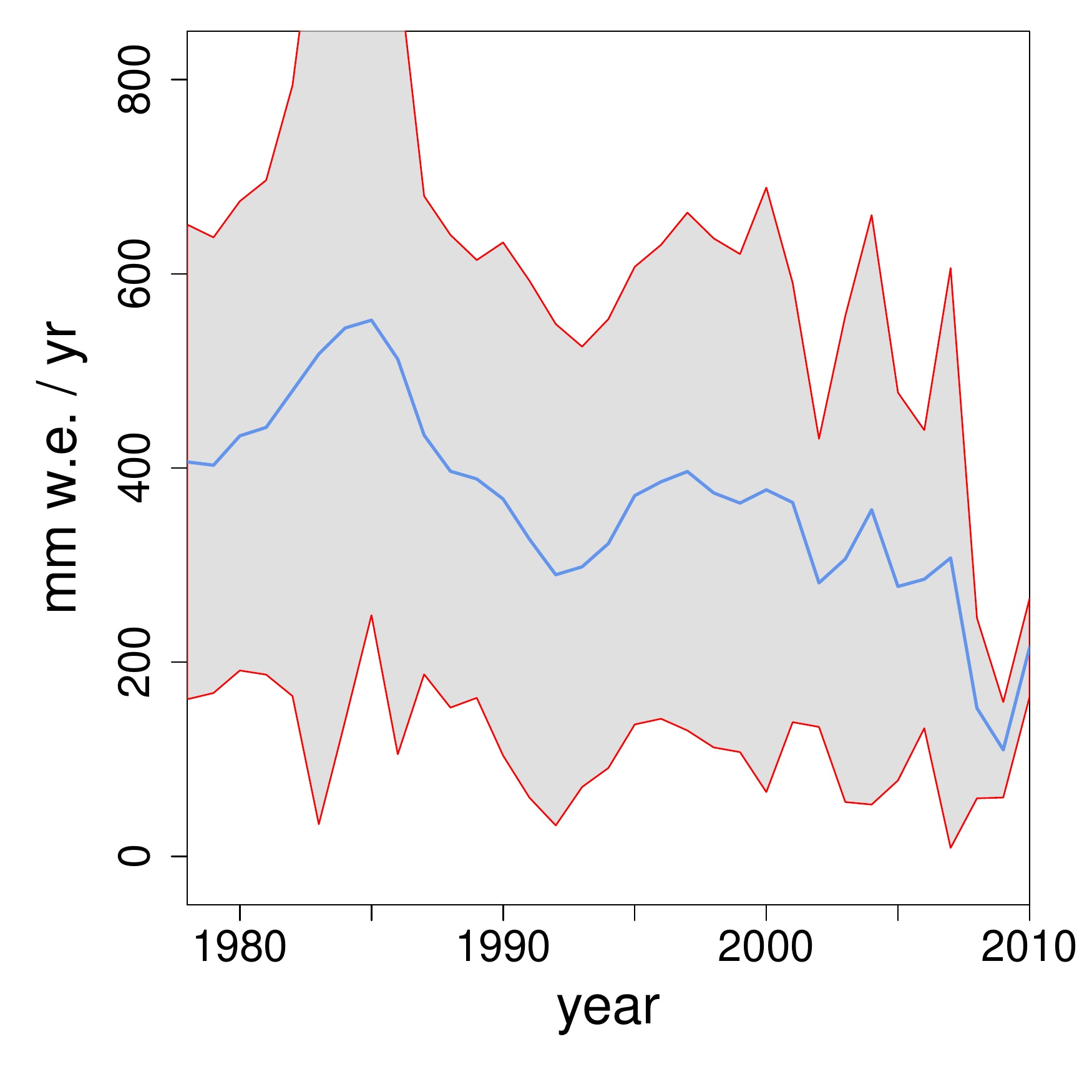}
        \includegraphics[width = 0.32 \textwidth]{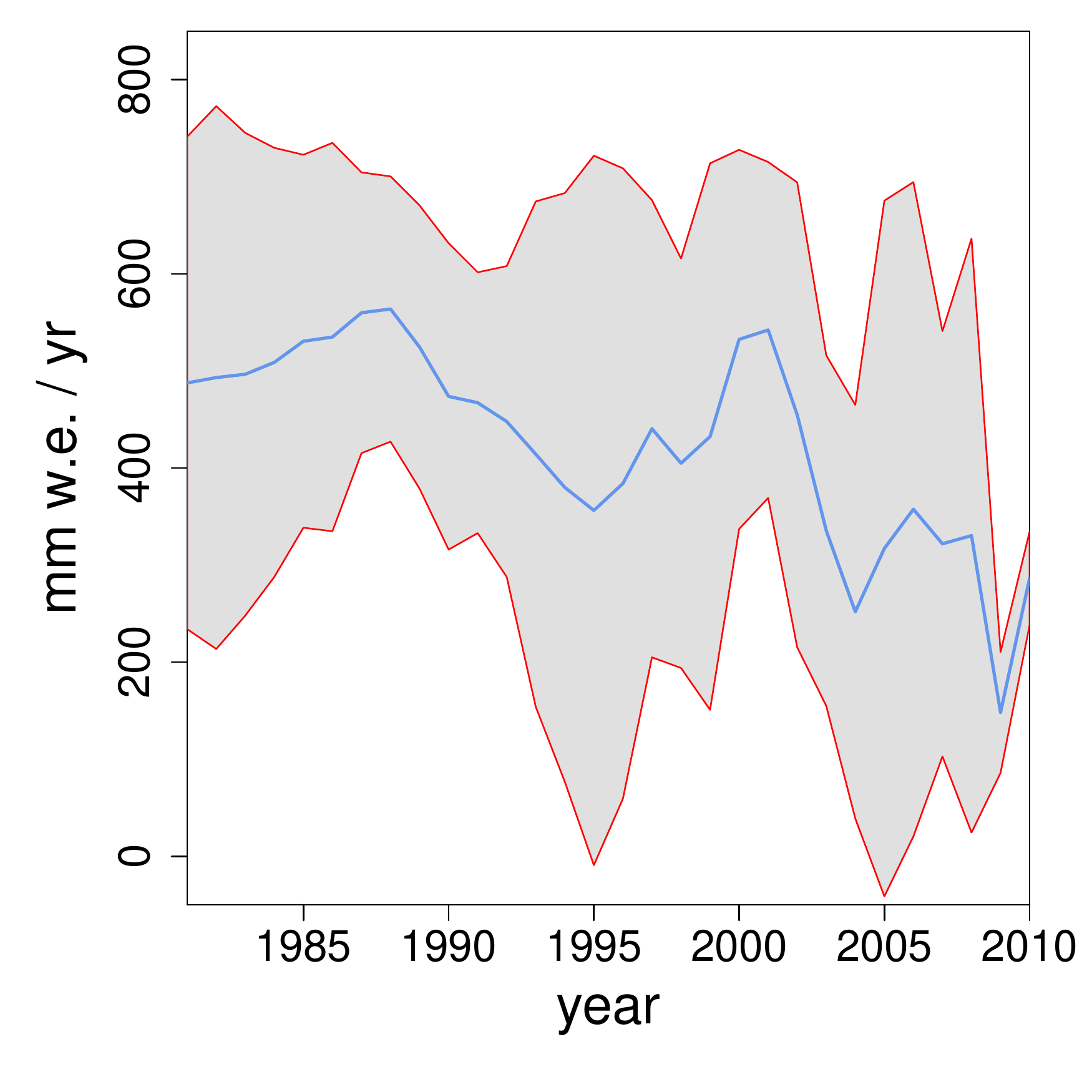}
             \includegraphics[width = 0.32 \textwidth]{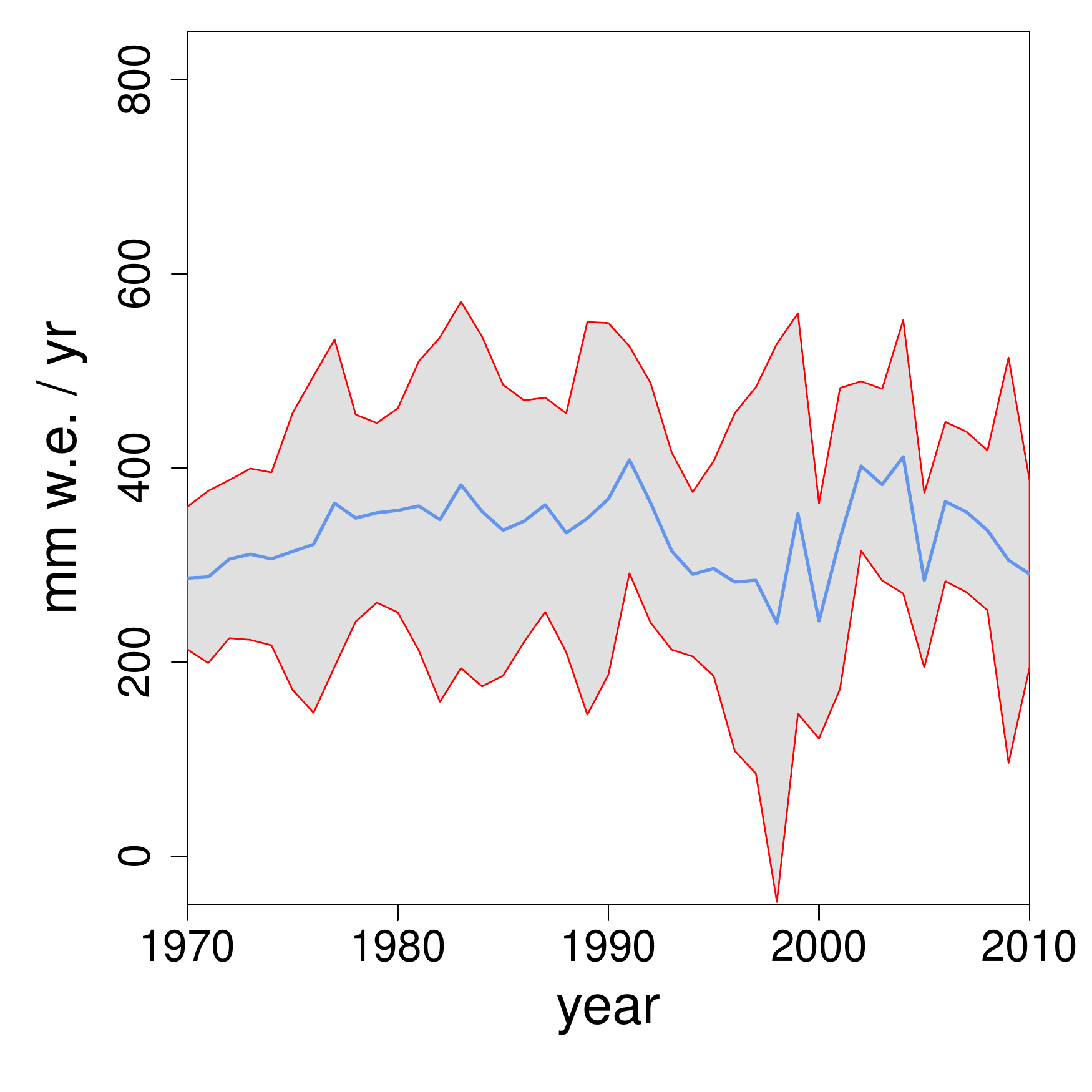}
                                     \vspace{-3mm}
    \caption{(Top) Estimates of snow density for sites A, B, and C, plotted in order left to right. (Bottom) Estimates of surface mass balance corresponding to the above estimates of snow density and ice-penetrating radar signals.}\label{fig:density_smb}
                            \vspace{-3mm}

\end{figure}

For the second task, we estimate snow density along much of the flight line where ice-penetrating radar measurements were taken (see Figures  \ref{fig:locs} and \ref{fig:echogram}). Due to low data quality at some locations, we were unable to obtain reliable SMB estimates. We estimate snow density down to 40 meters because, due to radar signal attenuation, these are the only densities useful for estimating SMB. Based on these snow density estimates along the flight line, we apply approaches from \cite{keeler2020}, briefly discussed in Section \ref{sec:firn}, to estimate SMB (with standard error) over recent decades. We focus on three locations without drilled snow cores, labeled ``A'', ``B'', and ``C'' in Figure \ref{fig:locs}, to illustrate how estimated snow densities can be leveraged to estimate SMB (See Figure \ref{fig:density_smb}). Sites A and B show two of the most negative SMB trends (-14.1 and -12.2 mm w.e./yr., on average), while site C has one of the most positive estimated SMB trends, 0.7 mm w.e./yr, on average.

\begin{figure}[ht]
                            \vspace{-3mm}

    \centering
        \includegraphics[width = 0.43 \textwidth]{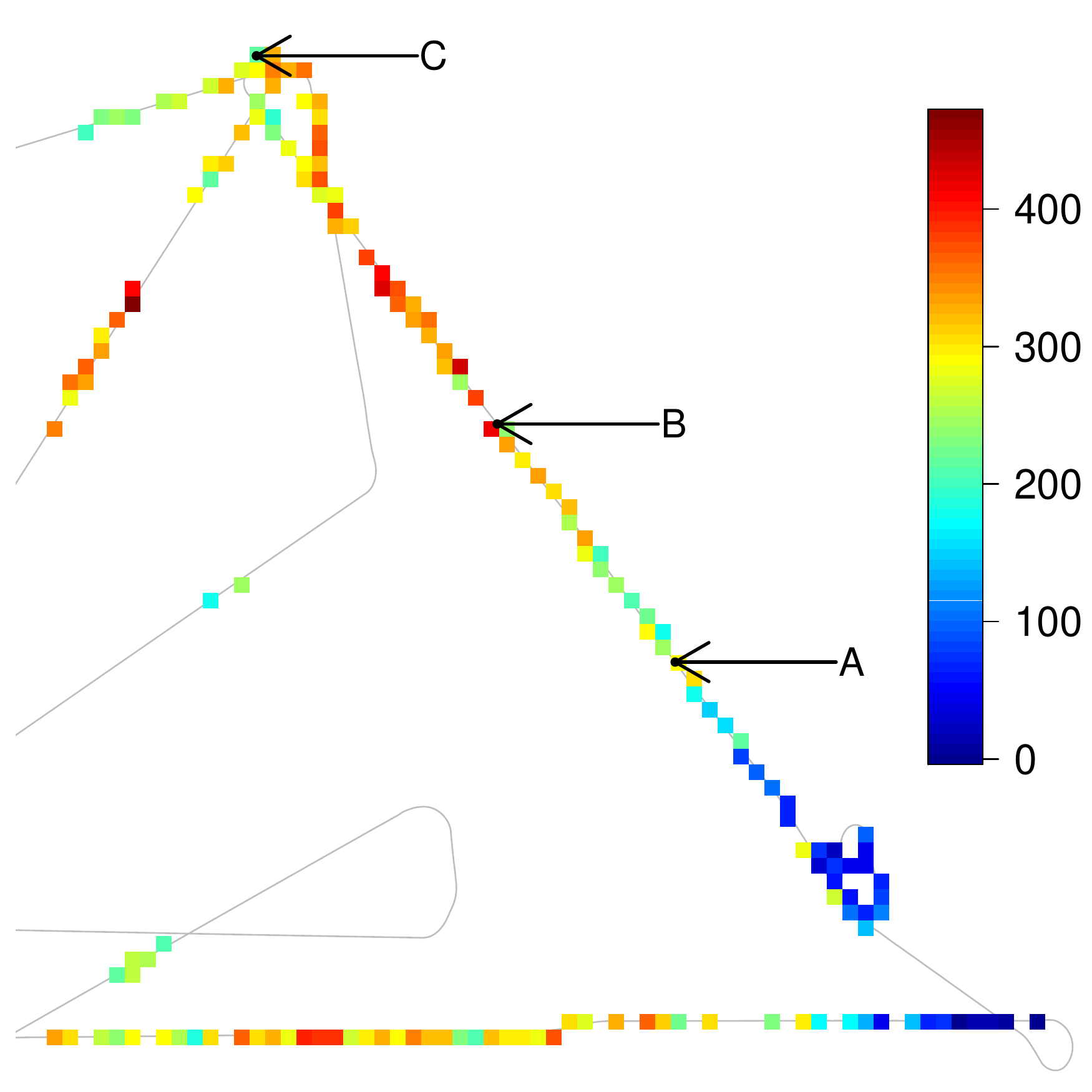}
        \hspace{3mm}
    \includegraphics[width = 0.53 \textwidth]{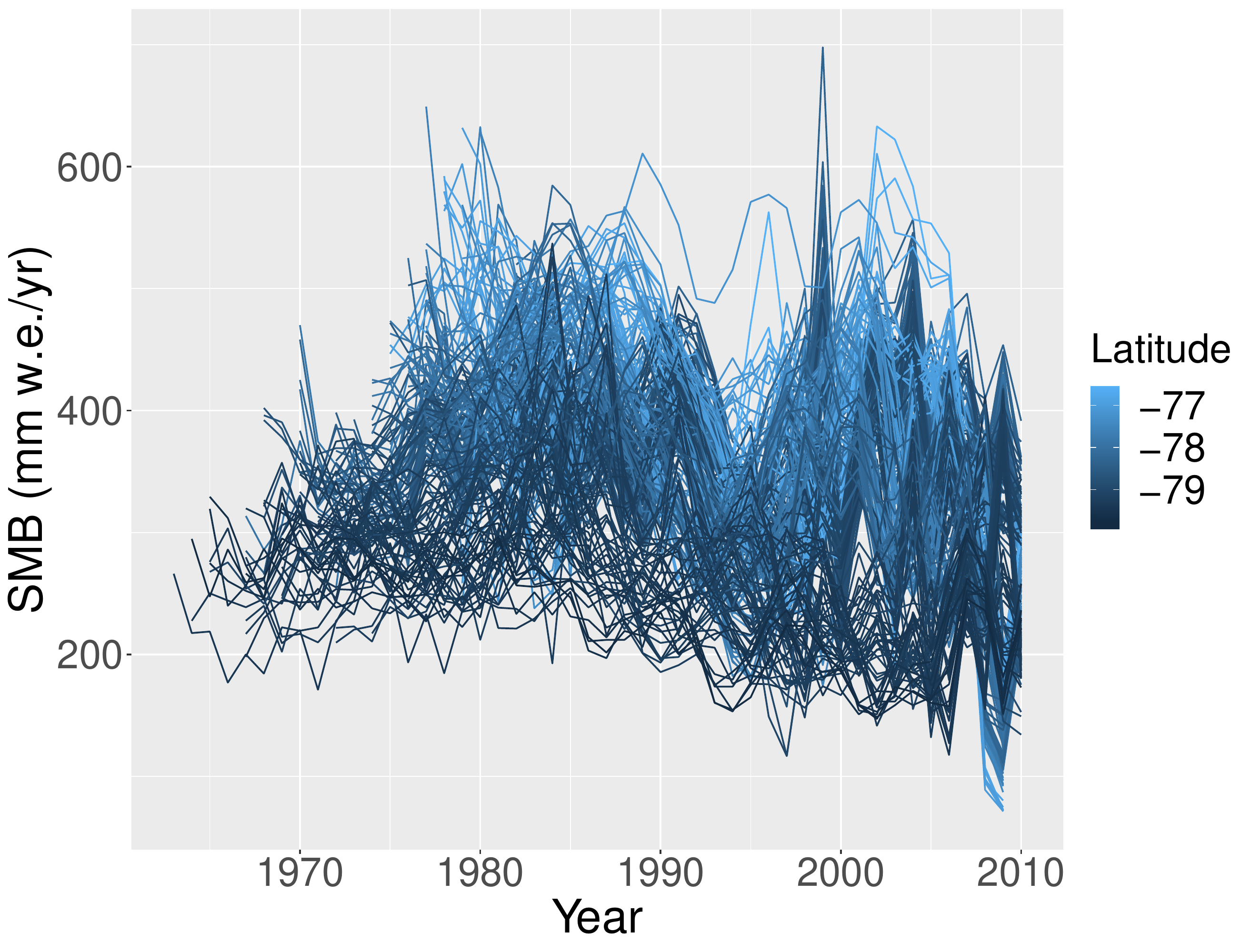}

    \caption{(Left) Estimated mean SMB (mm w.e./yr), averaged over all years (Right) Estimated mean SMB over all years, colored by latitude.}\label{fig:smb_res}
                            \vspace{-3mm}
\end{figure}

\begin{figure}[ht]
                            \vspace{-3mm}

    \centering
            \includegraphics[width = 0.43 \textwidth]{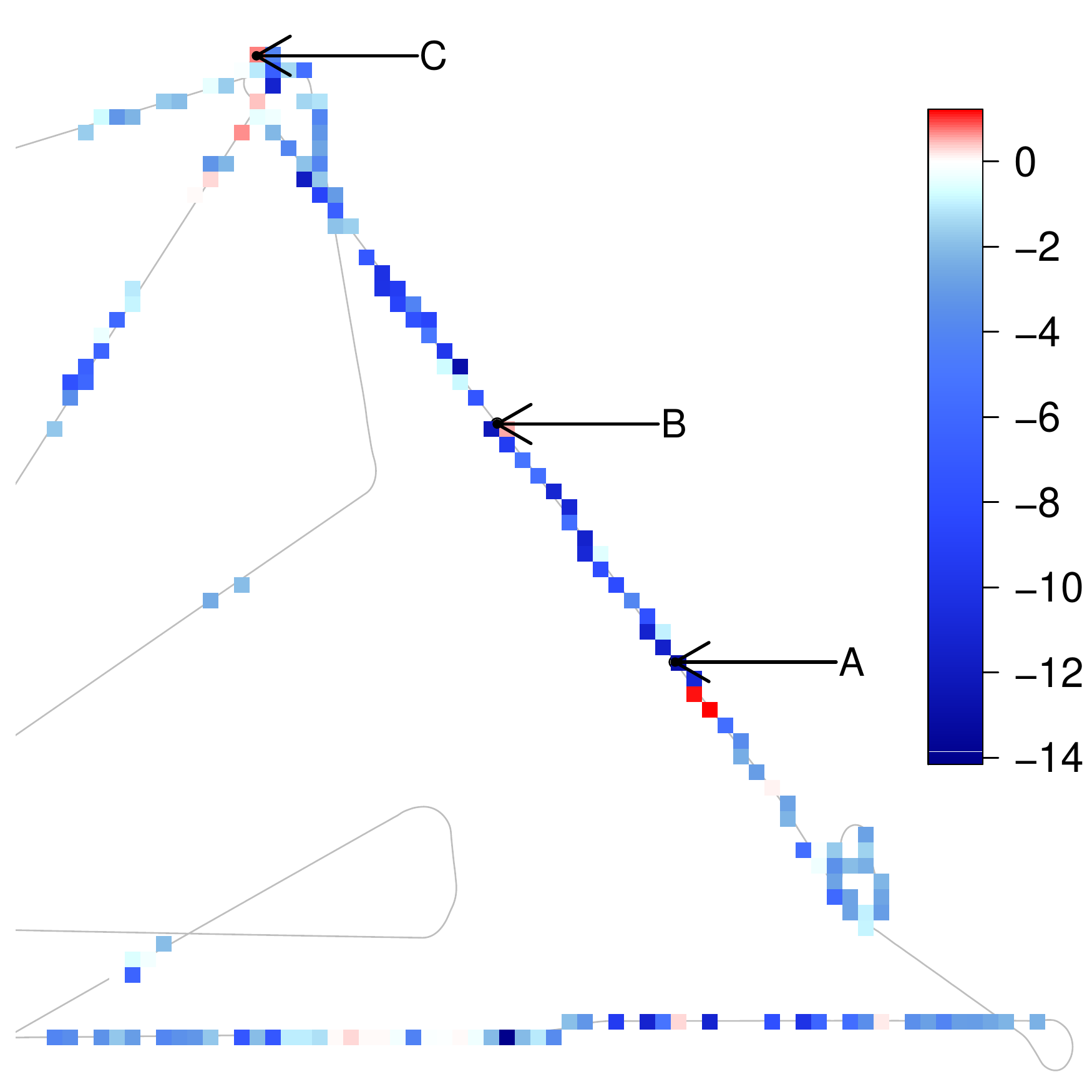}
    \includegraphics[width = 0.43 \textwidth]{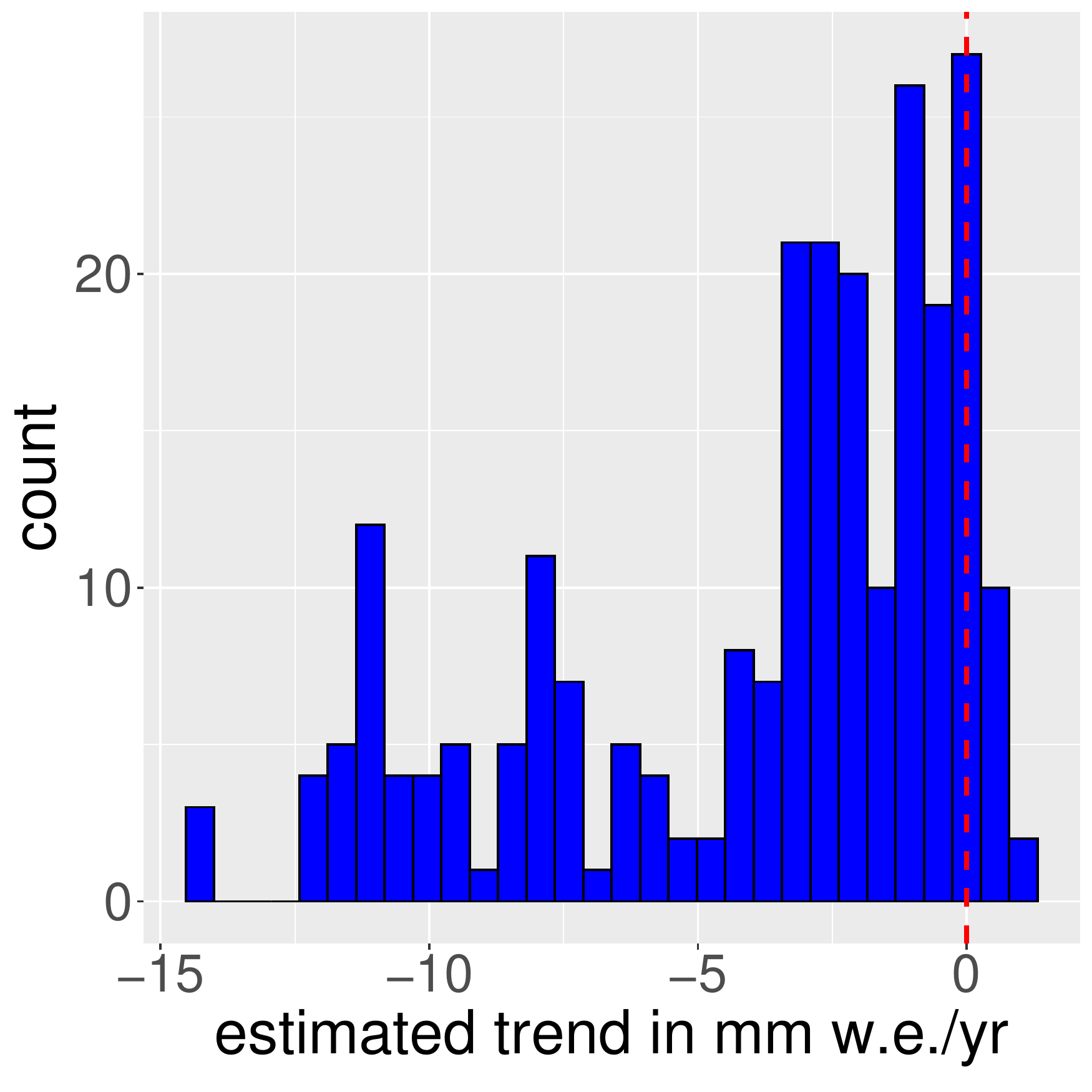}
    \caption{Estimated trend in mean SMB (mm w.e./yr) (Left) plotted by location and (Right) in histogram with red vertical line at 0.}\label{fig:smb_trends}
                            \vspace{-3mm}
\end{figure}

We plot the estimated SMB curves and time-averaged SMB estimates in Figure \ref{fig:smb_res} along the flight line. Overall, we see lower SMB at latitudes closer to the south pole. For each estimated SMB curve, we estimate the trend over time using a weighted linear model, using the inverse squared standard error as weights. We plot the estimated trends in SMB in a histogram and at their locations on the flight line in Figure \ref{fig:smb_trends}. Approximately 90\% of the sites along these flight lines have estimated negative slopes. Overall, these results suggest generally negative trends in SMB across the central West Antarctic ice sheet. 
The preponderance of negative SMB trends is particularly noteworthy as this region has experienced pronounced warming during the same time period \citep{bromwich_central_2013}.
As the moisture carrying capacity of air increases with temperature, the inverse relationship of SMB and air temperature in this region suggests changes in atmospheric moisture transport are the dominant driver of the observed SMB trends.

\section{Conclusions and Future Work}\label{sec:conc}

We have presented a novel class of spatial models for monotone curve estimation. For theses data, we constructed this model through kernel convolutions that enables prediction of snow density below observed depths and at locations where snow cores have not been drilled. Using 19-fold cross-validation, we selected a weighted spatially-varying generalized linear model using M-spline bases for the kernel convolution, where the coefficients of the integrated kernels are modeled using independent log-Gaussian processes. We have demonstrated how this model estimates snow density below observed levels and where no data has been acquired. Using these snow density estimates, we have estimated surface mass balance over the recent past and find generally negative trends in SMB over the West Antarctic Ice Sheets.

In this analysis, we did not account for the time of density measurement. In future work, we will account for time differences between the time of radar measurement and the snow core drilling through a spatiotemporal model. However, the demonstrated ability to interpolate snow densification rates has several applications relevant to the cryospheric community. 
As demonstrated, the extension of density profiles to arbitrary spatial locations is useful in estimating local annual SMB using ice-penetrating radar, greatly expanding the spatial density and coverage of these measurements over traditional coring techniques.
Similarly, quantifying the total mass loss in Antarctica and similar regions is a primary target within the community, with critical implications for sea-level rise \citep{ipcc_climate_2013}.
A primary method for determining net mass loss uses laser altimetry to measure changes in ice sheet surface elevation but also requires accurate estimates of snow density to estimate mass changes \citep{li_modeling_2011}.
The laser altimetry method, in particular, is a focus of several national and international laboratories and organizations, highlighted by NASA's recent ICESat-2 satellite launch in 2018 and the ongoing ESA CryoSat-2 laser altimeter mission \citep{tedesco_remote_2014}. Lastly, to improve snow density estimation, we will consider extensions of \cite{white2019}, where we consider sites that would sequentially minimize integrated error criteria over regions of Antarctica.

%\cite{meyer2008} shows that there are more general monotone piecewise cubic regression spline models than those available through I-spline basis functions. Therefore, possible extensions of this work could explore these more general models. These approaches, however, would likely need to leverage truncated multivariate normal models for regression coefficients that we found worse in out-of-sample prediction for this dataset than log-Gaussian models. In our model comparison, we found that a piecewise quadratic model with a very simple spatial model was preferable to more complicated cubic I-spline models. 

%We hypothesize that as more spatial sites are incorporated into this model (i.e., more firn cores become available) and we apply this approach to more heterogeneous areas, richer multivariate spatial models like the linear model of coregionalization or the Multivariate Mat\'ern covariance function \citep{gneiting2010} may become more competitive. Extensions of our modeling approach could also explore how climatic variables affect the rate of densification and corresponding coefficients in our kernel convolution model. In addition to possible to modeling advances, we are interested in spatial design for this setting. Specifically, we could consider extensions of \cite{white2019}, where we consider sites that would sequentially minimize integrated error criteria over regions of Antarctica.
A natural extension of our proposed framework could increase the number of knots included in the MISP and provide shrinkage on the integrated kernel coefficients through an autoregressive prior distribution \citep[see][]{lang2004bayesian}. This approach would likely reduce the model's sensitivity to knot selection. More importantly, this extension would enable the automation of this approach for other applications.

\section*{Acknowledgment}

We thank the three anonymous reviewers, as well as the Editor, for their comments that have helped improve the manuscript. Summer Rupper acknowledges funding from NASA grant NNX16AJ72G.
 
\appendix

\section{Covariance of MISP Model}

We recall the specification of the space-depth random effect $w(\bs,x)$, 
\begin{equation}
\begin{aligned}
w(\bs,x) &= \alpha(\bs) + \sum^J_{j = 1} K_j(x) z_j^*(\bs), \\ &= \alpha(\bs) + \bK(x)^T \bz^*(\bs),
\end{aligned}
\end{equation}
where $K_j(x) = \int_0^x k_j(u - \xi_j) \, du$, and $k_j(\cdot)$ and $z_j^*(\bs)$ are kernels and independent spatial log-Gaussian processes associated with each knot $\xi_j$. The vectors $\bK(x)$ and $\bz^*(\bs)$ contain $J$ elements of integrated kernels $K_j(x)$ and log-Gaussian process elements $z_j^*(\bs)$ for $j = 1,..,J$. We use independent GPs for $\alpha(\bs)$ and each $\log(z^*_j(\bs))$,
\begin{align}
\alpha(\bs) &\sim GP\left(\gamma_0,\sigma^2_0 e^{- \phi d(\bs,\bs')} \right) \\
\log(z^*_j(\bs)) &\overset{ind}{\sim} GP\left(\gamma_j, \sigma^2_j e^{ - \phi d(\bs,\bs') } \right),
\end{align}
for $j = 1,2,3,4,5,6$, and where $\sigma^2_{j}$ is the parameter-specific spatial variance, $d(\bs,\bs')$ is the great-circle distance between locations $\bs$ and $\bs'$, and $\phi$ is the common spatial covariance decay parameter for all parameters. 

Because we use a log-Gaussian process prior for $z_j^*(\bs)$, 
\begin{align*}
    \mathbb{E}\left(  z_j^*(\bs) \right) &=\exp\left[\mathbb{E}\left(  \log(z_j^*(\bs)) \right) +  \text{Var}(\log(z_j^*(\bs)) ) \right] =\exp\left(\gamma_j + \sigma^2_j/2\right) \\
    \text{Cov}\left(  z_j^*(\bs),z_j^*(\bs') \right)&= \mathbb{E}\left(  z_j^*(\bs) \right)  \mathbb{E}\left(  z_j^*(\bs') \right) \left[\exp\left(\text{Cov}(\log(z_j^*(\bs)),\log(z_j^*(\bs'))  ) \right) - 1 \right] \\
    &= \exp\left(2\gamma_j + \sigma^2_j\right)\left\lbrace\exp\left[\sigma^2_j \exp\left( - \phi d(\bs,\bs') \right)  \right] - 1\right\rbrace.
\end{align*}

To obtain the covariance of the final model, we insert these pieces into (3.3), giving $\text{Cov}\left(w(\bs,x), w(\bs',x') \right)$ as
\begin{equation}\label{eq:cov_final}
      \sigma^2_0 \exp\left( - \phi d(\bs,\bs') \right) + \sum_{j=1}^6 k_j(x) k_{j}(x') \exp\left(2\gamma_j + \sigma^2_{j} \right)\left\lbrace\exp\left[\sigma^2_j \exp\left( - \phi d(\bs,\bs')  \right)  \right] - 1\right\rbrace, 
\end{equation}
where $w(\bs,x)$ is the space-depth random function in Section 3.
Because we cannot generally express $k_j(x) k_j(x')$ as a function of the depth separation $|x - x'|$, this model is non-stationary. Moreover, this model is nonseparable as the covariance function cannot be factored into one function of the great-circle distance $d(\bs,\bs')$ and another function including only depth arguments ($x$ and $x'$). This model also allows the hierarchical mean $\gamma_j$ to influence the covariance of the model, giving an explicit relationship between the hierarchical mean and the model's covariance \citep{hefley2017basis}.

\section{Construction of M- and I-Spline Basis Functions}

Here, we define M-spline basis functions given a sequences of $L$ interior knots with a polynomial order $p$. Following \cite{ramsay1988}, conventionally, M-splines of order $l$ are defined such that it is a piecewise polynomial of order $p = l - 1$, and we follow this convention here. We choose $L$ ordered interior knots $x_{min} < \kappa_1<...<\kappa_L < x_{max}$, where $x_{min}$ and $x_{max}$ are the minimum and maximum observed depths, 0 and 140 m, respectively. 
 
We define an augmented knot sequence $\{ \xi_1,...,\xi_{L + 2l} \}$, defined as
\begin{itemize}
\item $x_{min} = \xi_1 = ... = \xi_l$,
\item $\xi_{l + j} = \kappa_{j}$, for $j = 1,...,L$, and
\item $x_{max} = \xi_{L + 1} = ... = \xi_{L + 2l}$.
\end{itemize}
These repeated boundary knots at $x_{min}$ and $x_{max}$ reduce boundary effects for higher-order ($l >1$) M-splines. Using this augmented knot sequence, \cite{ramsay1988} defines M-splines of order $l >1 $, a piecewise polynomial of order $p-1$, as 
$$ M_j^{(l)}(x) =\left\{\begin{array}{lr}
        \frac{l\left[ (x - \xi_j) M_j^{(l-1)}(x)  + (\xi_{j + l} - x) M_{j+1}^{(l-1)}(x) \right]}{(l-1)(\xi_{j + l} -  \xi_{j}) }, & \text{for } \xi_{j} \leq x\leq \xi_{j+l} \\
        0, & \text{otherwise }
        \end{array}\right\}, $$
for $j = 1,...,L + l$, where the first-order M-spline is 
$$ M_j^{(1)}(x) =\left\{\begin{array}{lr}
        \frac{1}{(\xi_{j + 1} -  \xi_{j}) }, & \text{for } \xi_{j} \leq x\leq \xi_{j+1} \\
        0, & \text{otherwise }
        \end{array}\right\} $$
for $j = 1,...,l + 1$.
This construction yields $L + l$ non-zero basis functions. If the coefficients of a linear combination of M-spline basis functions are non-negative, then the resulting function is positive. Because we introduce M-spline bases as a kernel for our kernel convolution of positive spatial processes, we highlight that M-spline bases are non-negative and integrate to 1; thus, they are probability densities. Moreover, $M_j^{(l)}(x)$ is zero unless $\xi_j \leq x < \xi_{j + l}$ so the linear model discussed in Section 3 is sparse, meaning that this model scales better for large datasets. To see more discussion on these and other properties, see \cite{ramsay1988} and \cite{meyer2008}.

Because our model integrates over smoothing kernels, we are interested in the I-spline basis function because they are defined as the integral of M-spline basis functions, 
\begin{equation}
    I_j^{(l)}(x) = \int^x_{\xi_1} M_j^{(l)}(t) dt, 
\end{equation}
for $j = 1,...,L + l$ and $x_{min} < x < x_{max}$. Because the I-spline basis functions are integrals of M-spline terms, if the M-spline model has polynomial degree $l-1$, then the I-spline is constructed of polynomials of degree $l$. If the coefficients of a linear combination of I-spline basis functions are non-negative, then the resulting function is monotone increasing.

To make this definition compatible with our definition of the MISP in Section 3, we note that there is an $M_j^{(l)}(\cdot)$ for each interior knot, as well as $l$ terms for $x_{min}$. The number of knots for the MISP is $J = L + l$, with $l$ knots at $x_{min}$ and one knot at each of the $L$ interior knots. As discussed in the manuscript, we consider $M_j^{(l)}(\cdot)$ is a suitable choice of $k_j(\cdot)$, and $I_j^{(l)}(\cdot)$ corresponds to the integrated kernel $K_j(\cdot)$.

\section{Models Comparison Results}\label{app:mod_comp}

In this section, we present model comparison results, focusing on specific components of the model in each subsection. For model comparison presentations, we fix all model components to those of the final model, except for the component we are focusing on. We use the integrated model comparison criteria presented in Section 4.1. When there is not agreement across all model comparison criteria, we use CRPS as our final model selection criterion because it is a proper scoring rule \citep{gneiting2007}.

\subsection{Comparison of Kernels}

In this subsection, we focus on the selection of the kernel for the kernel convolution of log-Gaussian spatial processes discussed in Section 3. We specifically consider Normal, $t$, Laplace, asymmetric Laplace, and M-spline kernels. For all of the results presented below, we use specifications for all other components corresponding to the final model. Specifically, we use campaign-specific variances with weighting proportional to the length of the core used to derive the measurement. In addition, we use independent Gaussian processes with shared decay parameters for $\alpha(\bs)$ and $\log(z_j^*(\bs))$. 

The results in Table \ref{tab:kernel_choice} represent a subset of many models considered, where each model is the best of the models considered for that kernel. In particular, we consider various configurations of the knots, as well as random/fixed kernel scale (bandwidth) parameters. We find that the M-spline basis is best in terms of all predictive criteria; however, the predictive performance is remarkably similar across kernel specifications. 

\begin{table}[H]
\centering
\scriptsize
\begin{tabular}{lrrrr}
  \hline
 Kernel &  ISE & IAE & CRPS & Relative CRPS \\ 
  \hline
Normal & 1.3820e-04 & 3.4818e-03 & 1.4867e-02 & 1.0843 \\ 
$t_4$& 1.4185e-04 & 3.4991e-03 & 1.4214e-02 & 1.0366 \\  
$t_{10}$ & 1.4042e-04 & 3.4852e-03 & 1.4173e-02 & 1.0336 \\  
Laplace & 1.5803e-04 & 3.6548e-03 & 1.4548e-02 & 1.0610 \\ 
Asymetric Laplace (left-skewed) & 1.4295e-04 & 3.5231e-03 & 1.3987e-02 & 1.0201 \\
Asymetric Laplace (right-skewed) &  1.4059e-04 & 3.4661e-03 & 1.4674e-02 & 1.0702  \\ 
M-spline &  1.3591e-04 & 3.4301e-03 & 1.3712e-02 & 1.0000 \\    \hline
\end{tabular}
\caption{Model comparison results for models differing by kernel.}\label{tab:kernel_choice}
\end{table}

\subsection{Comparison of Weighting and Error Models}

In this subsection, we focus on the selection of the weighting model for the truncated normal. We specifically consider homoscedastic, fixed weighting (as presented in the manuscript), as well as a random weighting $ (n_{\bs_i} / x_{max,\bs_i})^\eta $, where $\eta > 0$. For all of the results presented below, we use the M-spline basis kernel. We use independent log-Gaussian processes with shared decay parameters for $z_j^*(\bs)$. The results in Table \ref{tab:weight_choice} suggest that the best model in terms of CRPS uses fixed weights and campaign-specific variances. The best model here has a relative CRPS of 1. These results are expected since we use integrated quantities for model comparison (Section 4.1), and the fixed weights are equivalent to those in the approximated integrated quantities. 

\begin{table}[H]
\centering
\scriptsize
\begin{tabular}{llrrrr}
  \hline
Weighting & Campaign-Specific & ISE & IAE & CRPS & Relative CRPS \\ 
  \hline
Homoscedastic (none)& No & 1.3474e-04 & 3.4228e-03  & 1.4995e-02 & 1.0935 \\ 
Fixed Weighting & No & 1.4253e-04 & 3.6115e-03& 1.5787e-02 & 1.1513 \\
Fixed Weighting & Yes  & 1.3591e-04 & 3.4301e-03 & 1.3712e-02 & 1.0000  \\ 
Random Weighting & Yes & 1.3325e-04 & 3.3956e-03 & 1.4134e-02 & 1.0308 \\
  \hline

\end{tabular}
\caption{Model comparison results for models differing by weighting.}\label{tab:weight_choice}
\end{table}

We also include a comparison between the truncated Normal data model and the truncated $t$ with four degrees of freedom common to robust regression. We find that the truncated Normal model outperforms the $t$ distributed error model in terms of all criteria (See Table \ref{tab:error_choice}).

\begin{table}[H]
\centering
\scriptsize
\begin{tabular}{lrrrr}
  \hline
Data model &  ISE & IAE & CRPS & Relative CRPS \\ 
  \hline
Truncated Normal & 1.3591e-04 & 3.4301e-03 & 1.3712e-02 & 1.0000  \\ 
Truncated $t$ & 4.3157e-04 & 4.2074e-03 & 3.9525e-02 & 2.8825 \\ 
  \hline

\end{tabular}
\caption{Model comparison results for models differing by data model.}\label{tab:error_choice}
\end{table}

\subsection{M-spline Basis Specification}

In this subsection, we focus on the selection of the degree and knot locations of the M-spline kernels or bases. For all of the results presented below, we use the best selection of all other components of the model. Specifically, we use campaign-specific variances with weighting proportional to the length of the core used to derive the measurement. In addition, we use independent log-Gaussian processes with shared decay parameters for $z_j^*(\bs)$.  The results in Table \ref{tab:interior_choice} represent a subset of many models considered. In particular, for each number of interior knots, we used various knot locations and present those with the best predictive performance. For the best model, we present a comparison of knot locations in Table \ref{tab:knot_loc_choice}. We choose the model with the lowest CRPS but note that it is not lowest in ISE and IAE; however, it is very close. The best model here has a relative CRPS of 1. The final model uses piecewise constant kernels on the generalized logit-scale like many firn density models, including \citep{herron_firn_1980} and \cite{verjans2020bayesian}. This model, however, has more discontinuities than previous models and is a spatially-varying model that allows spatial interpolation of firn density curves. Interestingly, higher-order I-spline models were better using different weighting schemes; however, these were not as good as the best model presented here.

\begin{table}[H]
\centering
\scriptsize
\begin{tabular}{rrrrrr}
  \hline
I-spline Degree & Interior Knots & ISE & IAE & CRPS & Relative CRPS \\ 
  \hline
  1 & 3 & 1.3983e-04 & 3.4170e-03 & 1.4024e-02 & 1.0227 \\ 
  2 & 3 & 1.3581e-04 & 3.3866e-03 & 1.4308e-02 & 1.0435 \\ 
  3 & 3 & 1.3914e-04 & 3.4198e-03 & 1.4935e-02 & 1.0892 \\ 
  1 & 5 & 1.3591e-04 & 3.4301e-03 & 1.3712e-02 & 1.0000  \\ 
  2 & 5 & 1.8374e-04 & 3.9003e-03 & 1.7499e-02 & 1.2762 \\
  3 & 5 & 2.4493e-04 & 4.8249e-03 & 2.3457e-02 & 1.7107 \\ 
  1 & 7 & 1.3481e-04 & 3.3827e-03 & 1.3893e-02 & 1.0132 \\
  2 & 7 & 2.3783e-04 & 4.8034e-03 & 2.3783e-02 & 1.7345 \\ 
  3 & 7 & 2.0563e-04 & 4.4703e-03 & 2.4953e-02 & 1.8198 \\ 
     \hline
\end{tabular}
\caption{Model comparison results for models differing by the number of interior knots and polynomial degree of the integrated M-spline (I-spline) basis.}\label{tab:interior_choice}
\end{table}

\begin{table}[ht]
\centering
\scriptsize
\begin{tabular}{rlrrrr}
  \hline
Selection & Knot Locations (in m) &  ISE & IAE & CRPS & Relative CRPS \\ 
  \hline
1  & 0.0, 3.1, 6.50, 9.8, 13.2, 17.1 & 1.3471e-04 & 3.4278e-03 & 1.4184e-02 & 1.0344 \\ 
2 & 0.0, 1.8, 4.8, 9.8, 16, 53.6 & 1.3625e-04 & 3.4132e-03 & 1.3841e-02 & 1.0094 \\ 
3 & 0.0, 23.3, 46.7, 70, 93.3, 116.7 & 1.5830e-04 & 3.9557e-03 & 1.6695e-02 & 1.2175 \\ 
4 & 0.0, 5.0, 15.0, 30.0, 45.0, 70.0 &  1.3591e-04 & 3.4301e-03 & 1.3712e-02 & 1.0000  \\ 
5 & 0.0, 2.0, 5.0, 10.0, 20.0, 45.0 &  1.3605e-04 & 3.3861e-03& 1.3866e-02 & 1.0112 \\ 
   \hline
\end{tabular}
\caption{Model comparison results for models differing by knot locations. We present five selections: 1. evenly-spaced depth quantiles, 2. unevenly spaced quantiles, 3. evenly-spaced, 4. unevenly-spaced with more flexibility at shallow depths, 5. unevenly-spaced with more knots at shallow depths.}\label{tab:knot_loc_choice}
\end{table}

\subsection{Spatial Model for Log-Gaussian Processes}

We now present a comparison of some of the spatial models considered for the intercept and log-Gaussian processes used in the process convolution model. For all models here, we use the truncated Normal data model with a campaign-specific depth-weighted variance parameter. We also use M-spline kernels which yield and I-spline model when integrated. We find the best covariance specification for the corresponding spatial terms (the intercept $\alpha(\bs)$ and $\log(z_j^*(\bs))$) are specified by independent GP models with exponential covariance with shared spatial decay parameter $\phi$ (See Table \ref{tab:cross-covariance}). Also, each Gaussian process prior is centered on a unique scalar mean. We consider joint specifications using multivariate Gaussian processes for $\alpha(\bs)$ and $\log(z_j^*(\bs))$ but ultimately find that the simple independent GPs for these terms with a single decay parameter outperforms the more complex models (See Table \ref{tab:cross-covariance}).

\begin{table}[H]
\centering
\begin{tabular}{lrrrr}
  \hline
Spatial Model & ISE & IAE & CRPS & Relative CRPS \\ 
  \hline
Independent GP, one $\phi$ & \textbf{1.3591e-04} & \textbf{3.4301e-03} & 1.3712e-02 & 1.0000  \\ 
Independent GP & 1.3901e-04 & 3.5508e-03 & 1.4038e-02 & 1.0238 \\ 
Separable & 1.6073e-04 & 3.7400e-03 & 1.4554e-02 & 1.0614 \\ 
Coregionalization & 1.4185e-03 & 1.2656e-02 & 3.9132e-02 & 2.8539 \\ 
   \hline
\end{tabular}
\caption{Model comparison results for joint covariance specification of $\alpha(\bs)$ and $\log(z_j^*(\bs))$. }\label{tab:cross-covariance}
\end{table}

We also consider including a linear mixed model that adds variance to the diagonal of covariances models for $\alpha(\bs)$ and $\log(z_j^*(\bs))$. Ultimately, however, the mixed effect model was worse in terms of predictive performance (See Table \ref{tab:mixed_effect}). 

\begin{table}[H]
\caption{Model comparison between purely spatial covariance model and mixed effects specification models for spatial models.}\label{tab:mixed_effect}
\centering
\begin{tabular}{lrrrr}
  \hline
Model & ISE & IAE & CRPS & Relative CRPS \\ 
  \hline
Purely Spatial & \textbf{1.3591e-04} & \textbf{3.4301e-03} & 1.3712e-02 & 1.0000 \\ 
Mixed Effect & 1.5209e-04 & 3.8186e-03  & 1.6206e-02 & 1.1819 \\ 
   \hline
\end{tabular}

\end{table}

The last comparison we present is for the marginal covariance functions for $\alpha(\bs)$ and $\log(z_j^*(\bs))$ (See Table \ref{tab:matern}). We consider four specifications from the Mat\'ern covariance class. Using the great-circle distance, we only use the exponential covariance (the Mat\'ern class with smoothness $\nu = 1/2$) because the Mat\'ern class is not positive-definite when $\nu > 1/2$ using the the great-circle distance. The Mat\'ern class using spherical Euclidean distance is positive definite for all $\nu$; therefore, we consider this distance metric in combination with $\nu = 1/2, 3/2, 5/2$. We find that the Mat\'ern covariance with $\nu = 1/2$ is best in terms of prediction for these data. The smoother Mat\'ern covariance functions ($\nu = 3/2,5/2$) posed computational instability when $\phi$ was small because some sites are as close as 158 meters, and these covariance models yielded significantly worse predictions than the exponential covariance function. We also find a small advantage in using the great-circle distance compared to the three-dimensional Euclidean distance but wish to emphasize that these differences are small.

\begin{table}[H]
\caption{Model comparison between Mat\'ern covariance specifications.}\label{tab:matern}
\centering
\begin{tabular}{llrrrr}
  \hline
Distance & Smoothness $\nu$ & ISE & IAE & CRPS & Relative CRPS \\ 
  \hline
Great Circle & 1/2 & \textbf{1.3591e-04} & \textbf{3.4301e-03} & 1.3712e-02 & 1.0000 \\ 
3-D Euclidean & 1/2  & 1.3901e-04 & 3.4354e-03 & 1.3757e-02 & 1.0032 \\ 
3-D Euclidean & 3/2 & 2.1895e-04 & 4.5882e-03 & 1.5673e-02 & 1.1430 \\   
3-D Euclidean & 5/2 & 3.4429e-04 & 5.7260e-03 & 2.0946e-02 & 1.5276 \\ 
   \hline
\end{tabular}

\end{table}

We hypothesize that as more spatial sites are incorporated into this model (i.e., more firn cores become available) and we apply this approach to more heterogeneous areas or applications, richer multivariate spatial models like the linear model of coregionalization or the Multivariate Mat\'ern covariance function \citep{gneiting2010} may become more competitive. Extensions of our modeling approach could also explore how climatic variables affect the rate of densification and corresponding coefficients in our kernel convolution model. When applied to new regions or applications with more spatial heterogeneity, non-stationary models may also be beneficial.

\section{MCMC Model Fitting Results and Diagnostics}\label{app:MCMC}

\subsection{Extended Posterior Analysis}

Using 50,000 posterior samples, we provide the posterior summaries (posterior mean, standard deviation, 2.5\%, and 97.5\%) for all parameters except spatially-distributed intercepts and I-spline coefficients in Table \ref{tab:post_sum}. Interestingly, the posterior means of $\gamma_1$, $\gamma_2$, and $\gamma_3$ are approximately equal; however, there is significant variability in the corresponding site-specific parameters ($z_1^*(\bs)$, $z_2^*(\bs)$, $z_3^*(\bs)$). Likewise, the posterior means of $\gamma_5$ and $\gamma_6$ are the same, but the site-specific parameters are not. Because the knots are not evenly spaced over depth, this does not imply that the densification rate is not constant. Instead, this means that these segments contribute equally to $\log( \mu(\bs,x) / (\rho_I - \mu(\bs,x)))$. The campaign-specific variances differ significantly, suggesting that these campaigns contributed different levels of noise to the data. 

\begin{table}[ht]
\centering
\footnotesize
        \vspace{-4mm}
\caption{Posterior means, standard deviations, 2.5\%, and 97.5\% for non-spatial parameters.}\label{tab:post_sum}
\begin{tabular}{rrrrr}
  \hline
 & mean & sd & 2.5\% & 97.5\% \\ 
  \hline
$\tau^2_\text{EAP}$ & 7.81e-05 & 7.92e-06 & 6.41e-05 & 9.51e-05 \\ 
$\tau^2_\text{SDM}$ & 1.63e-04 & 1.12e-05 & 1.43e-04 & 1.87e-04 \\ 
$\tau^2_\text{SEAT}$ & 1.08e-05 & 1.39e-07 & 1.05e-05 & 1.11e-05 \\ 
$\tau^2_\text{US}$ & 1.90e-04 & 6.20e-06 & 1.78e-04 & 2.03e-04 \\
$\phi$ & 3.30e-04 & 7.59e-05 & 2.05e-04 & 5.00e-04 \\
$\sigma^2_0$ & 2.63e-01 & 6.74e-02 & 1.61e-01 & 4.21e-01 \\ 
$\sigma^2_1$ & 4.58e-01 & 1.36e-01 & 2.56e-01 & 7.82e-01 \\ 
$\sigma^2_2$ & 6.72e-01 & 2.05e-01 & 3.73e-01 & 1.16e+00 \\  
$\sigma^2_3$ & 5.95e-01 & 1.97e-01 & 3.14e-01 & 1.08e+00 \\ 
$\sigma^2_4$ & 1.14e+00 & 3.89e-01 & 5.82e-01 & 2.08e+00 \\
$\sigma^2_5$ & 8.83e-01 & 3.10e-01 & 4.49e-01 & 1.65e+00 \\
$\sigma^2_6$ & 5.29e+00 & 2.73e+00 & 1.95e+00 & 1.22e+01 \\
$\gamma_0$ & -3.78e-01 & 3.86e-01 & -1.15e+00 & 3.79e-01 \\ 
$e^{\gamma_1}$ & 3.89e-01 & 2.00e-01 & 1.30e-01 & 8.82e-01 \\
$e^{\gamma_2}$ & 3.87e-01 & 2.33e-01 & 1.09e-01 & 9.84e-01 \\
$e^{\gamma_3}$ & 3.94e-01 & 2.29e-01 & 1.15e-01 & 9.77e-01 \\
$e^{\gamma_4}$ & 3.57e-01 & 2.62e-01 & 7.77e-02 & 1.03e+00 \\ 
$e^{\gamma_5}$ & 5.50e-01 & 3.71e-01 & 1.29e-01 & 1.50e+00 \\ 
$e^{\gamma_6}$ & 5.52e-01 & 5.91e-01 & 6.40e-02 & 2.09e+00 \\
   \hline
\end{tabular}

        \vspace{-2mm}

\end{table}

\subsection{MCMC Convergence Diagnostics}

Although many parameters can be updated in closed form, we use Hamiltonian Monte Carlo, implemented using the Stan programming language \citep{carpenter2017}. Here, we provide convergence diagnostics for this model. 

\begin{figure}[ht]
\begin{center}
\includegraphics[width=0.45\textwidth]{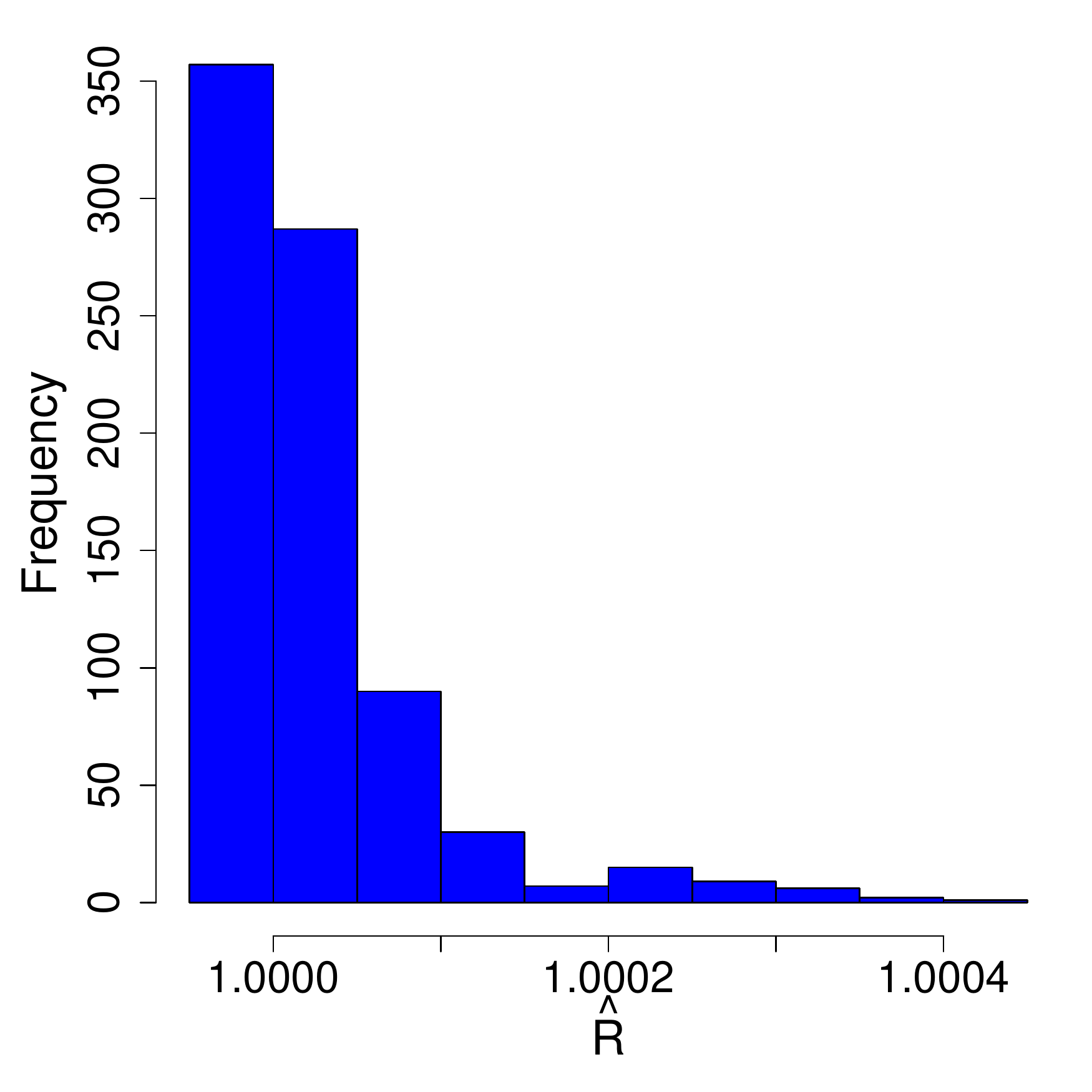}
\includegraphics[width=0.45\textwidth]{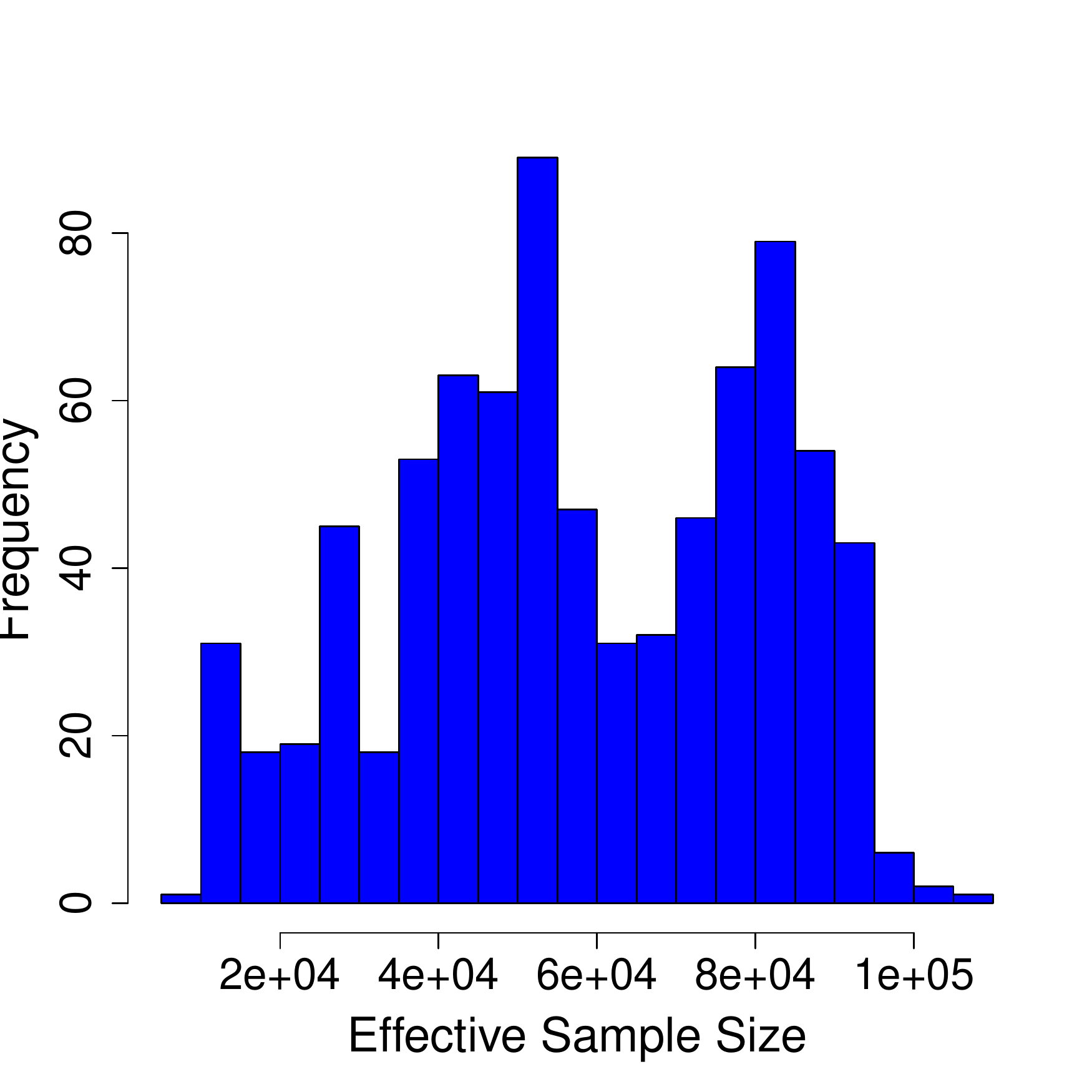}

\end{center}
\caption{Histogram of (Left) $\hat{R}$ for all model parameters using four MCMC chains and (Right) effective samples sizes.}\label{fig:Rhat_hist}
\end{figure}

First, we use the updated multi-chain diagnostic $\hat{R}$, proposed by \cite{vehtari2020rank} as an improvement over \cite{gelman1992inference}. We fit our final model four times, each with 25,000 post-burn iterations. As a heuristic, \cite{vehtari2020rank} suggest $\hat{R} > 1.01$ signal convergence issues. In our model fitting, all parameters have $\hat{R} < 1.0005$ (See Figure \ref{fig:Rhat_hist}), signaling each chain is sampling from the same distribution. 

For our final MCMC run, we obtain 50,000 post-burn samples and compute the effective sample size (ESS) \citep{geyer1992} of each parameter. The minimum ESS is 7113 for the spatial scale parameter $\sigma^2_6$. All other parameters have an ESS greater than 10,000.

\begin{figure}[H]
    \centering
    \includegraphics[width = 0.4\textwidth]{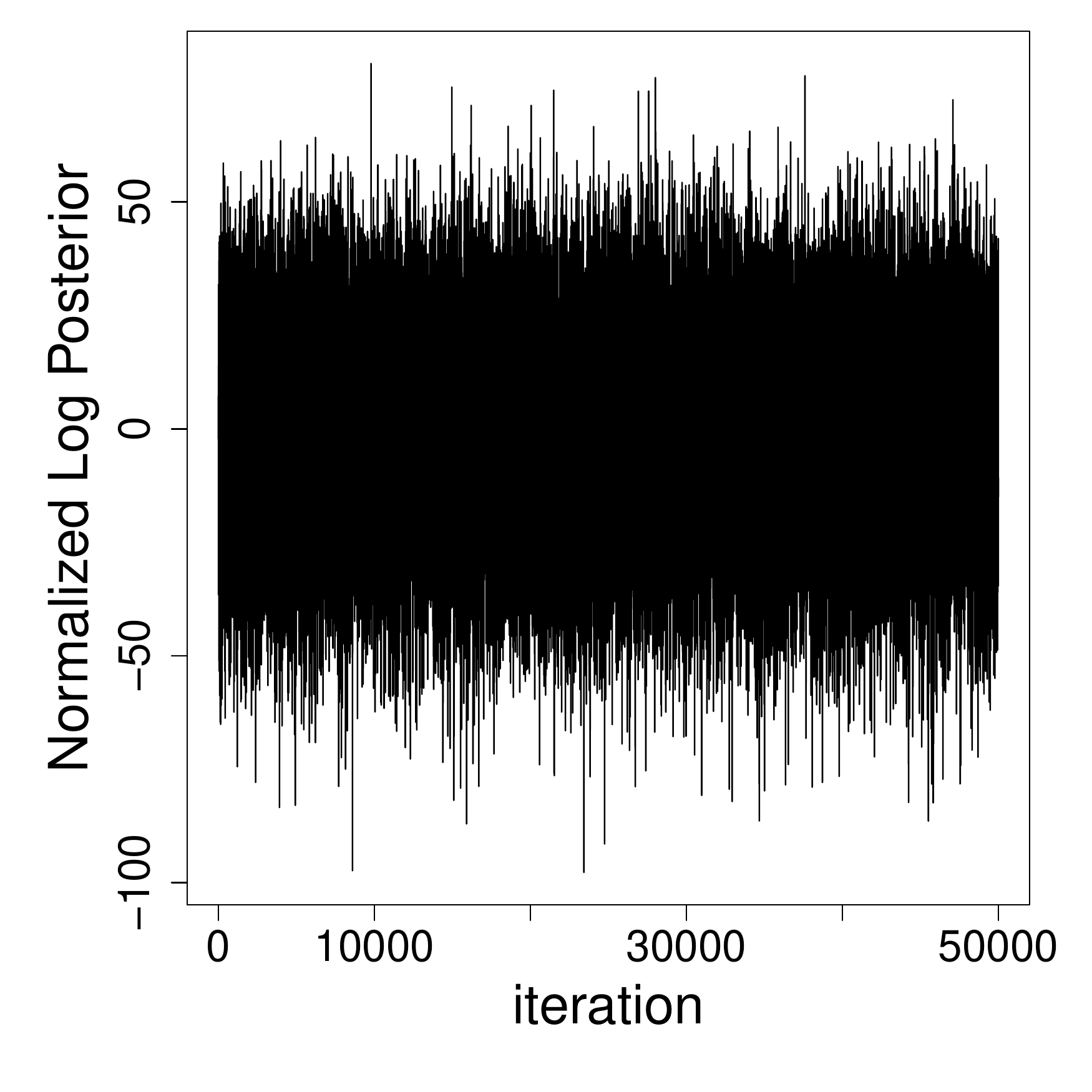}
    \caption{Trace plot of the normalized (to have mean zero) log posterior.}
    \label{fig:log_lik_trace}
\end{figure}

\begin{figure}[H]

    \centering
    \includegraphics[width = 0.7\textwidth]{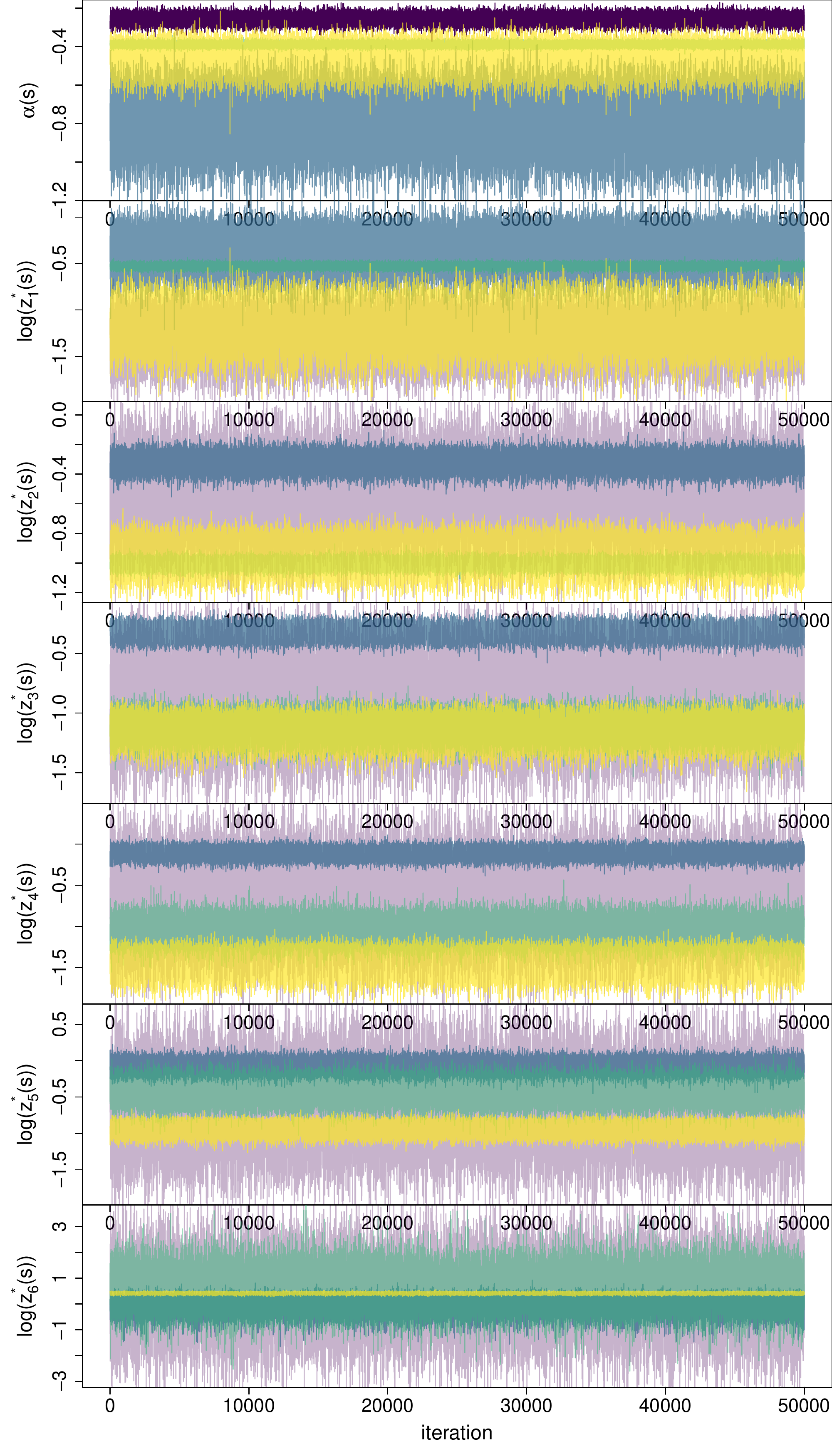} 
        \caption{Trace Plots for spatially-varying intercepts and log-regression coefficients. Samples are plotted in the following colors: core 1 in Purple, core 8 in blue, core 21 in green, and core 43 in yellow.}    \label{fig:trace_plots}
\end{figure}

For spatial generalized linear models, the mixing of spatial parameters is often of greatest concern. To illustrate good mixing, in Figure \ref{fig:trace_plots}, we plot trace plots for the spatial parameters at sites 1, 8, 21, 42 (cores 1,8, 21, and 43 due to one replicate). These cores are 2, 100, 18, and 140 meters long, respectively. These sites were chosen because these cores come from different campaigns and are spatially spread out. All trace plots are well-behaved. We also plot the scaled log-posterior in Figure \ref{fig:log_lik_trace} to show good mixing of the model as a whole.

\bibliographystyle{apacite}
\bibliography{ref}

\end{document}